\documentclass[preprint,showpacs,preprintnumbers,amsmath,amssymb]{revtex4}
    \usepackage{amsmath}
    \usepackage{amsfonts}
    \usepackage{graphicx}
    \usepackage{dcolumn}
    \usepackage{bm}
    \begin{document}

    \title{Longitudinal polarization of hyperon and anti-hyperon in semi-inclusive deep-inelastic scattering}
    \author {Shan-shan Zhou, Ye Chen, Zuo-tang Liang and Qing-hua Xu}
    \affiliation{School of Physics, Shandong University, Jinan, Shandong 250100, China}

    \date{\today}

    \vspace*{0.3cm}

    \begin{abstract}
    We make a detailed study of the longitudinal polarization of
    hyperons and anti-hyperons in semi-inclusive deep-inelastic
    lepton-nucleon scattering. We present the numerical results for spin
    transfer in quark fragmentation processes, analyze the possible
    origins for a difference between the polarization for hyperon and
    that for the corresponding anti-hyperon.
    We present the results obtained in the case that there is no asymmetry
    between sea and anti-sea distribution in nucleon as well as
    those obtained when such an asymmetry is taken into account.
    We compare the results with the
    available data such as those from COMPASS and make predictions for
    future experiments including those at even higher energies such as at eRHIC.
    \end{abstract}
    \maketitle

    \section{Introduction}

    Because of the non-perturbative nature, our knowledge on hadron
    structure and that on the fragmentation function are still very much
    limited, in particular in the polarized case. Deeply inelastic lepton-nucleon
    scattering is always an ideal place for such study because at
    sufficiently high energy and momentum transfer, factorization
    theorem is applicable and the hard part is easy to be calculated.
    Hyperon polarizations have been widely used for such studies, since
    they can easily be determined by measuring the angular distributions
    of the decay products. These studies have attracted much attention
    in last years.~[see e.g., \cite{Buskulic:1996vb,Ackerstaff:1997nh,Astier:2000ax,Adams:1999px,
    Airapetian:1999sh,Alexakhin:2005dz,Sapozhnikov:2005sc,Xu:2005js,
    Gustafson:1992iq, Boros:1998kc,
    Liu:2001yt,Liu:2000fi,Xu:2002hz,Zuotang:2002ub,Dong:2005ea,Xu:2004es,Xu:2005ru,Chen:2007tm,
    Ellis:1995fc,Kotzinian:1997vd, Ellis:2002zv,Ellis:2007ig,Ma:1998pd,Ma:2000cg,
    Ma:2000uu,Ma:2000uv,Anselmino:2000ga,Anselmino:2001ps}.]  
    Longitudinal polarizations of hyperons and
    anti-hyperons in semi-inclusive deep-inelastic scattering (SIDIS)
    have been studied both experimentally and theoretically. 
    More recently, such studies have in particular been extended to anti-hyperons.
    Special attention is paid to the comparison of the results for hyperons with
    those for the corresponding anti-hyperons. This is partly triggered
    by the results of COMPASS collaboration at CERN which seem to tell
    us that there is a difference between $\Lambda$ and
    $\bar{\Lambda}$ polarization in semi-inclusive deep-inelastic
    lepton-nucleon scattering~\cite{Alexakhin:2005dz,Sapozhnikov:2005sc}. 
    A detailed study of such a difference can provide us
    useful information on the polarized fragmentation function and the
    structure of nucleon sea. It might be
    considered as a signature of the existence of a difference between
    the strange sea and anti-sea distributions in nucleon as proposed in literature some time 
    ago~\cite{Signal:1987gz,Brodsky:1996hc,Burkardt:1991di,Holtmann:1996be,
    Christiansen:1998dz,Cao:1999da,Olness:2003wz,Lai:2007dq}. 
    It could also be a signature for
    a difference between the spin transfer in quark and anti-quark fragmentation.
    On the other hand, it is also clear that the valence quarks in nucleon and
    other known effects can also contribute to such a difference. 
    It is therefore important to make a detailed and
    systematic analysis of the contributions from such known effects
    before we extract information on the possible asymmetry between sea
    and anti-sea distributions.

    In this paper, we make such a systematic study of longitudinal
    polarization of different hyperons and anti-hyperons in
    semi-inclusive deep-inelastic lepton-nucleon scattering. We make a detailed
    analysis on the possible origin(s) of the difference between hyperon
    and anti-hyperon polarization at COMPASS and even higher energies.
    We clarify the different contributions and present the results obtained in the case that there is no asymmetry
    between nucleon sea and anti-sea quark distributions as well as
    those obtained when such an asymmetry is taken into account.
    We make the calculations not only for $\Lambda$ and $\bar{\Lambda}$ but also
    other hyperons and anti-hyperons in the same $J^P=(1/2)^+$ octet.
    We compare our results with the available data and make predictions for
    future experiments in particular at eRHIC.\cite{Deshpande:2005wd}

    The paper is organized as follows: After this introduction, in
    Sec.  II, we summarize the general framework for the calculations
    of the longitudinal polarization $P_H$ of the hyperon $H$ and
    $P_{\bar H}$ of the anti-hyperon $\bar H$ based on factorization theorem,
    and make a detailed analysis of each
    factor used in the formulae. We present in particular the model calculation results
    for spin transfer for a pure quark fragmentation process and compare the
    results for hyperons with those for anti-hyperons.
    In Sec. III, we present the results obtained for hyperon and anti-hyperon polarizations in
    reactions using polarized beam and unpolarized target for the case
    that there is no asymmetry between the sea and anti-sea
    distributions in nucleon and those for the case that such an asymmetry
    is taken into account. We also study the influence from the
    differences in quark distributions as given by different
    sets of parameterizations. 
    In Sec. IV, we study the case that
    the lepton beam is unpolarized but the nucleon is polarized and
    present the results obtained using different parameterizations of the
    polarized parton distributions. 
    Finally, in Sec. V, we give a
    short summary and discussion.

    \section{General framework for calculating $P_H$ and $P_{\bar H}$ in SIDIS}
    Deeply inelastic lepton-nucleon scattering at sufficiently high
    energy and momentum transfer is one of the places where
    factorization theorem is applicable and is tested with high
    accuracies. According to the factorization theorem, hadron
    production in the current fragmentation region of SIDIS is a
    pure result of the fragmentation of the quark or anti-quark
    scattered by the incoming lepton. The cross section is given as a
    convolution of quark distribution function in nucleon, the
    elementary lepton-parton scattering and the fragmentation function.
    We consider the longitudinally polarized reaction in this paper and
    for definiteness, we consider $e^-+N\to e^-+H ({\rm or}\ \bar H)+X$ as
    an example. The formulae can be extended to other reactions in a
    straight forward way. To the leading order in perturbation theory,
    the differential cross section for
    $e^-+N\to e^-+H+X$ is given by,
    \begin{equation}
    d\sigma_{\lambda_H;\lambda_e,\lambda_N}=\sum_{f,\lambda_f}\int dx dy dz K(x,y)
    \left[ q_{f,\lambda_f}^{N,\lambda_N}(x)d\hat{\sigma}^{eq}_{\lambda_e,\lambda_f}(x,y)
    D_{f,\lambda_f}^{H,\lambda_H}(z)
    +(q_f\leftrightarrow \bar q_f)\right],
    \label{eq:factorization}
    \end{equation}
    where $\lambda_e$, $\lambda_N$, $\lambda_f$ and $\lambda_H$ are
    respectively the helicities of the electron, the incoming nucleon,
    the struck quark $q_f$ and the produced hyperon $H$; $x$ is the
    usual Bjorken-$x$, $y$ is the fractional energy transfer from the
    electron to the nucleon $N$ in the rest frame of $N$; $z$ is the
    fraction of momentum of scattered $q_f$ carried by the produced
    hyperon $H$; and $K(x,y)$ is a kinematic factor which contains the
    $1/Q^4$ due to the photon propagator and others ($Q^2=-q^2$ and $q$
    is the four momentum transfer). The sum over $f$ runs over all the
    different flavor of quarks or anti-quarks. Here, for clarity, we did
    not write out the scale dependence of the parton distributions and
    fragmentation functions explicitly. They are understood implicitly.
    We consider only light quarks and anti-quarks. Hence both quark and
    electron mass are neglected so that helicity in the elementary
    scattering process $eq\to eq$ is conserved.

    Eq.~(\ref{eq:factorization}) is the basis for calculating the cross
    section of SIDIS both in unpolarized and polarized case. We use
    this formulae as the starting point for calculating the
    polarizations of hyperons and anti-hyperons in SIDIS in the
    following but discuss possible violation effects in Sec. IIF.

    \subsection{The calculation formulae for $P_H$ and $P_{\bar H}$}

    The polarization of $H$ in $e^-+N\to e^-+H ({\rm or}\ \bar H)+X$ is usually defined as,
    \begin{equation}
    P_H(z)\equiv\frac{d\sigma_{+;\lambda_e\lambda_N}-d\sigma_{-;\lambda_e\lambda_N}}
    {d\sigma_{+;\lambda_e\lambda_N}+d\sigma_{-;\lambda_e\lambda_N}},
    \label{eq:definition}
    \end{equation}
    for the case that both the beam and target are completely polarized in the
    pure states with helicities $\lambda_e$ and $\lambda_N$.
    In the case that factorization theorem is valid, we can just insert
    Eq.~(\ref{eq:factorization}) into Eq.~(\ref{eq:definition}) and
    obtain the result for the polarization of hyperon in $e^-+p\to
    e^-+H+X$ with longitudinally polarized electron beam and proton
    target as,
    \begin{equation}
    P_{H}(z)=\frac{\sum_{f}e^{2}_{f}\int dxdy K(x,y)
    \left\{P_{f}(x,y)\left[q_{f}(x)+P_bP_TD_L(y)\Delta q_f(x)\right]\Delta{D^{H}_f(z)}+(q_f\leftrightarrow\bar q_f)\right\}}
    {\sum_fe^2_f\int dxdy K(x,y)
    \left\{\left[q_f(x)+P_bP_TD_L(y)\Delta q_f(x)\right]D^H_f(z)+(q_f\leftrightarrow\bar q_f) \right\} },
    \label{eq:polH}
    \end{equation}
    where $P_b$ and $P_T$ denote the longitudinal polarization of the
    electron beam and nucleon target respectively;
    $e_f$ is the electric charge of quark $q_f$, $q_f(x)$ and $\Delta q_f(x)$
    are the unpolarized and polarized quark distribution functions,
    $P_f(x,y)$ is the polarization of the scattered quark $q_f$, $D_L(y)$ is the longitudinal spin transfer factor
    in the elementary scattering process $eq\to eq$ and is defined as,
    \begin{equation}
    D_L(y)\equiv \frac{d\hat\sigma^{eq}_{++}-d\hat\sigma^{eq}_{+-}}{d\hat\sigma^{eq}_{++}+d\hat\sigma^{eq}_{+-}},
    \label{eq:DLY1}
    \end{equation}
    which is only a function of $y$ at the leading order in perturbative QED;
    $D^H_f(z)$ and $\Delta D^H_f(z)$ are the unpolarized and
    polarized fragmentation functions that are defined as,
    \begin{equation}
    D^{H}_f(z)\equiv D^{H}_f(z,+)+D^{H}_f(z,-),
    \label{eq:DfH}
    \end{equation}
    \begin{equation}
    \Delta{D^{H}_f(z)}\equiv D^{H}_f(z,+)-D^{H}_f(z,-),
    \label{eq:deltaDfH}
    \end{equation}
    where the argument $+$ or $-$ denotes that the helicity of the
    produced hyperon $H$ is the same as or opposite to that of the
    fragmenting $q_f$. In the notation used in
    Eq.~(\ref{eq:factorization}),  they are
    $D_f^H(z,+)=D_{f,+}^{H,+}(z)=D_{f,-}^{H,-}(z)$ and
    $D_f^H(z,-)=D_{f,-}^{H,+}(z)=D_{f,+}^{H,-}(z)$.
    The integrations over $x$ and $y$ run over the kinematic region determined by
    the corresponding experiments.

    Similarly for anti-hyperon $\bar H$ in $e^-+p\to e^-+\bar H+X$, the polarization is given by,
    \begin{equation}
    P_{\bar{H}}(z)=\frac{\sum_{f}e^{2}_{f}\int dxdyK(x,y)
    \left\{P_{f}(x,y)\left[q_{f}(x)+P_bP_TD_L(y)\Delta q_f(x)\right]\Delta{D^{\bar H}_f(z)}+(q_f\leftrightarrow\bar q_f)\right\}}
    {\sum_fe^2_f\int dxdyK(x,y)
    \left\{\left[q_f(x)+P_bP_TD_L(y)\Delta q_f(x)\right]D^{\bar H}_f(z)+(q_f\leftrightarrow\bar q_f) \right\}},
    \label{eq:polHbar}
    \end{equation}

    The physical significance of the expressions in Eqs.~(\ref{eq:polH})
    and~(\ref{eq:polHbar}) are very clear: In the denominator, besides
    some kinematic factor, we have just the production rate of $H$ or
    $\bar H$. The appearance of the term proportional to $P_bP_T$ is due
    to the double spin asymmetry $\hat a_{LL}$ in the elementary
    scattering process $eq\to eq$ which measures the difference between
    $\hat\sigma^{eq}_{++}$ and $\hat\sigma^{eq}_{+-}$. The numerator
    shows explicitly that the polarization of $H$ or $\bar H$ just comes
    from that of the $q_f$ and/or $\bar q_f$ after the $eq$ scattering.
    This can be seen more clearly if we re-write Eqs.~(\ref{eq:polH}) and
    ~(\ref{eq:polHbar}) as,
    \begin{equation}
    P_H(z)=\sum_f\int dxdy\left[P_f(x,y)R_f^H(x,y,z|pol)S_f^H(z)+(q_f\leftrightarrow\bar q_f)\right],
    \label{eq:polHsimp}
    \end{equation}
    \begin{equation}
    P_{\bar H}(z)=\sum_f\int dxdy\left[P_f(x,y)R_f^{\bar H}(x,y,z|pol)S_f^{\bar H}(z)+(q_f\leftrightarrow\bar q_f)\right],
    \label{eq:polHbarsimp}
    \end{equation}
    where $R_f^H(x,y,z|pol)$ is the fractional contribution from $q_f$ to the production of $H$ in $e^-+p\to e^-+H+X$
    and is given by,
    \begin{equation}
    R_f^H(x,y,z|pol)=\frac{e_f^2K(x,y)\left[q_f(x)+P_bP_TD_L(y)\Delta q_f(x)\right]D^H_f(z)}
    {\sum_fe_f^2\int dxdyK(x,y)\left\{\left[q_f(x)+P_bP_TD_L(y)\Delta q_f(x)\right]D^H_f(z)+(q_f\leftrightarrow\bar q_f)\right\}};
    \label{eq:RfHpol}
    \end{equation}
    $S_f^H(z)$ is the polarization transfer in the fragmentation process $q_f\to H+X$ in the longitudinally polarized case 
    and is defined as,
    \begin{equation}
    S_f^H(z)\equiv \Delta D_f^H(z)/D_f^H(z).
    \label{eq:Stransfer}
    \end{equation}
    We see that the polarization of $H$ or $\bar H$, $P_H(z)$ or
    $P_{\bar H}(z)$, is just a weighted sum of $S_f^H(z)$ and $S_{\bar f}^H(z)$
    for different flavor $f$.
    The weights are products of $P_f(x,y)$, the polarization of quark after the elementary
    scattering, and $R_f^H(x,y,z|pol)$, the fractional contribution from
    $q_f$ to the production of $H$. In fact, assuming the validity of
    factorization theorem, fragmentation functions should be universal
    so that $S_f^H(z)$ and $S_{\bar f}^H(z)$ are also universal.
    Different results for $P_H(z)$ in different kinematic regions and/or
    different reactions just originate from the differences in
    $P_f(x,y)$ and $R_f^H(x,y,z|pol)$.

    The expression for the relative weight $R_f^H(x,y,z|pol)$ is much simpler if we have only
    beam or target polarized, i.e., we have either $P_T=0$ or $P_b=0$.
    In this case, the term proportional to $P_bP_T$ vanishes and we
    have, $R_f^H(x,y,z|pol)|_{P_b=0}=R_f^H(x,y,z|pol)|_{P_T=0}=R_f^H(x,y,z|unpol)$,
    which we simply denote by $R_f^H(x,y,z)$ and is given by,
    \begin{equation}
    R_f^H(x,y,z)=\frac{e_f^2K(x,y)q_f(x)D^H_f(z)}
    {\sum_fe_f^2\int dxdy K(x,y)\left[q_f(x)D^H_f(z)+\bar q_f(x)D^H_{\bar f}(z)\right]}.
    \label{eq:Rf|Pi=0}
    \end{equation}
    We see that $R_f^H(x,y,z)$ is determined solely
    by the unpolarized quantities such as the unpolarized parton
    distributions and fragmentation functions.

    The quark polarization is determined by the initial quark and/or electron polarization and the
    spin transfer in the elementary process. It is given by,
    \begin{equation}
    P_f(x,y)=\frac{P_bD_L(y)q_f(x)+P_T\Delta q_f(x)}
    {q_f(x)+P_bD_L(y)P_T\Delta q_f(x)},
    \label{eq:Pf}
    \end{equation}
    \begin{equation}
    P_{\bar f}(x,y)=\frac{P_bD_L(y)\bar q_f(x)+P_T\Delta \bar q_f(x) }
    {\bar q_f(x)+P_bD_L(y)P_T\Delta \bar q_f(x)},
    \label{eq:Pfbar}
    \end{equation}
    where the longitudinal spin transfer factor $D_L(y)$ in $eq\to eq$
    can be obtained using perturbative QED and, to the leading order, is
    the same for $eq\to eq$ and $e\bar q\to e\bar q$ and is given by,
    \begin{equation}
    D_L(y)=\frac{1-(1-y)^2}{1-(1+y)^2}.
    \label{eq:DLY2}
    \end{equation}

    This result has the following features.

    (1) If the target is unpolarized, i.e. $P_T=0$ but $P_b\not =0$, we have,
    \begin{equation}
    P_f(x,y|P_T=0)=P_{\bar f}(x,y|P_T=0)=P_bD_L(y),
    \label{eq:Pf|PT=0}
    \end{equation}
    which is only a function of $y$ and is the same not only for
    different flavors but also for quark and anti-quark.
    We see that, the quark (anti-quark) polarization in this case is completely
    known. This is a very good place to study the spin transfer in
    fragmentation and/or the factors contained in the fractional
    contributions to the production of $H$ and $\bar H$. The expression
    for hyperon polarization in this case becomes also simpler.
    It is given by,
    \begin{equation}
    P_{H}(z|P_T=0)=\int dxdy P_{b}D_{L}(y)\sum_{f} \left[R_f^H(x,y,z)S_f^H(z)+R_{\bar f}^H(x,y,z)S_{\bar f}^H(z)\right],
    \label{eq:polH|PT=0}
    \end{equation}
    \begin{equation}
    P_{\bar H}(z|P_T=0)=\int dxdy P_{b}D_{L}(y)\sum_{f}
    \left[R_f^{\bar H}(x,y,z)S_f^{\bar H}(z)+R_{\bar f}^{\bar H}(x,y,z)S_{\bar f}^{\bar H}(z)\right].
    \label{eq:polHbar|PT=0}
    \end{equation}
    For a fixed value of $y$, we have,
    \begin{equation}
    P_{H}(z,y|P_T=0)=\int dx P_{b}D_{L}(y)\sum_{f}
    \left[R_{fy}^H(x,y,z)S_f^{H}(z)+R_{\bar fy}^H(x,y,z)S_{\bar f}^{H}(z)\right],
    \end{equation}
    \begin{equation}
    R_{fy}^H(x,y,z)=\frac{e_f^2K(x,y)q_f(x)D^H_f(z)}
    {\sum_fe_f^2\int dx K(x,y)\left[q_f(x)D^H_f(z)+\bar q_f(x)D^H_{\bar f}(z)\right]}.
    \label{eq:Rfy}
    \end{equation}
    If we now define $S_{ep}^H(z,y)\equiv P_H(z,y|P_T=0)/P_bD_L(y)$ 
    as in COMPASS measurements\cite{Alexakhin:2005dz,Sapozhnikov:2005sc}, 
    we obtain that,
    \begin{equation}
    S_{ep}^{H}(z,y)= \sum_f  \int dx \left[ R_{fy}^H(x,y,z) S_f^H(z)+R_{\bar fy}^H(x,y,z)S_{\bar f}^H(z)\right].
    \label{eq:epspintransfer1}
    \end{equation}
    We see that $S_{ep}^{H}(z,y)$ is just a weighted sum of $S_f^H(z)$ and $S_{\bar f}^H(z)$,
    and the weights are determined by unpolarized quantities.
    Denote,
    \begin{equation}
    \langle R_{fy}^H(y,z)\rangle \equiv \int dx R_{fy}^H(x,y,z) =
    \frac{e_f^2\int dx K(x,y)q_f(x)D^H_f(z)}
    {\sum_fe_f^2\int dx K(x,y)\left[q_f(x)D^H_f(z)+\bar q_f(x)D^H_{\bar f}(z)\right]},
    \end{equation}
    and we obtain,
    \begin{equation}
    S_{ep}^{H}(z,y)= \sum_f  \left[ \langle R_{fy}^H(y,z)\rangle S_f^H(z)+\langle R_{\bar fy}^H(y,z)\rangle S_{\bar f}^H(z)\right].
    \label{eq:epspintransfer2}
    \end{equation}
    In practice, one often deals with events in a given $y$ interval, and one has, 
    \begin{equation}
    S_{ep}^{H}(z)= \sum_f  \left[ \langle R_{fyint}^H(z)\rangle S_f^H(z)+\langle R_{\bar fyint}^H(z)\rangle S_{\bar f}^H(z)\right],
    \label{eq:epspintransfer3}
    \end{equation}
    \begin{equation}
    \langle R_{fyint}^H(z)\rangle =
    \frac{e_f^2\int dxdy K(x,y)q_f(x)D^H_f(z)}
    {\sum_fe_f^2\int dxdy K(x,y)\left[q_f(x)D^H_f(z)+\bar q_f(x)D^H_{\bar f}(z)\right]}.
    \end{equation}
    
    (2) If the beam is unpolarized but the target is polarized, i.e.  $P_b=0$ but $P_T\not =0$, we have,
    \begin{eqnarray}
    &P_f(x,y|P_b=0)=P_T\Delta q_f(x)/q_f(x),\\
    &P_{\bar f}(x,y|P_b=0)=P_T\Delta \bar q_f(x)/\bar q_f(x),
    \label{eq:qqbarpol}
    \end{eqnarray}
    which is nothing else but the quark polarization before the $eq$ scattering. This is just a result of helicity conservation.
    In this case, we have,
    \begin{equation}
    P_H(z)=\frac{\sum_fe^2_fP_T\left[ \Delta q_f(x)\Delta D^H_f(z)+\Delta \bar q_f(x)\Delta D^H_{\bar f}(z) \right]}
    {\sum_fe^2_f\left[ q_f(x) D^H_f(z)+ \bar q_f(x) D^H_{\bar f}(z) \right]},
    \label{eq:polH|Pb=0}
    \end{equation}
    where the polarizations of hyperons and anti-hyperons are determined by the polarizations
    of quarks and anti-quarks in nucleon, thus can be used to extract information on the polarized
    quark distributions in nucleon.

    (3) In the case that neither $P_b$ nor $P_T$ is zero, i.e. both electron beam and nucleon target are polarized,
    the polarization of  $q_f$ or $\bar q_f$  after the scattering with electron is mainly
    determined by the beam electron polarization.
    It is dominated by the spin transfer from the electron to the scattered quark (anti-quark).
    The influence from the target polarization is relatively small.
    Aiming at studying either fragmentation functions or quark distributions, this case does not have much advantage
    compared to the cases (1) and (2) mentioned above.
    We therefore concentrate on the cases (1) and (2) in the following of this paper.

    From the discussions presented above, we see that there are three factors,
    i.e., quark polarization, the relative weights, and the fragmentation function are involved in
    Eqs.~(\ref{eq:polHsimp}-\ref{eq:Stransfer})
    for final hyperon or anti-hyperon polarization.
    We now discuss them further separately in the following.

    \subsection{The spin transfer factor $D_L(y)$ in $eq\to eq$ scattering}
    This is one of the best known factors involved in Eqs.~(\ref{eq:polH}) and (\ref{eq:polHbar}). 
    Since it is determined mainly by the electromagnetic interaction, the spin
    transfer factor $D_L(y)$ in $eq\to eq$ scattering is calculable
    using perturbation theory in QED. When next leading order effects
    are taken into account, pQCD corrections are involved. The result
    given in Eq.~(\ref{eq:DLY2}) is obtained at leading order in perturbation theory.
    In this case, $D_L(y)$ is the same for quark and anti-quark, i.e.,
    \begin{equation}
    D_L^{eq\to eq}(y)=D_L^{e\bar q\to e\bar q}(y).
    \end{equation}
    It is also the same for electron or positron. However, if we
    consider next-leading order in QED, e.g., if we take two photon
    exchange into account, the interference term leads to a difference
    between quark and anti-quark. It is also obvious that next leading
    order in QED is far away from influencing the results at the
    accuracies of the data available.
    We consider only the leading order here.

    Similarly, we also stick to leading order in perturbative QCD.
    The next-to-leading order calculations are in principle
    straight-forward but much involved (see e.g. \cite{Jager:2002xm}). 
    These results should
    be used consistently with the polarized parton distributions
    functions and the polarized fragmentation functions. In view of our
    current knowledge on the polarized fragmentation functions,
    we consider only the leading order consistently in this paper.

    \subsection{The relative weights and the parton distributions}
    Using charge conjugation symmetry for the fragmentation functions, we have,
    \begin{equation}
    R_{\bar f}^{\bar H}(x,y,z)=\frac{e_f^2K(x,y)\bar q_f(x)D^H_f(z)}
    {\sum_fe_f^2\int dxdy K(x,y)\left[\bar q_f(x)D^H_f(z)+q_f(x)D^H_{\bar f}(z)\right]}.
    \label{eq:Rfbar|Pi=0}
    \end{equation}
    This is to compare with $R_f^H(x,y,z)$ given in Eq.~(\ref{eq:Rf|Pi=0}).
    We see that the difference between $q_f(x)$ and $\bar q_f(x)$ is the only source for
    the difference between $R_f^H(x,y,z)$ and $R_{\bar f}^{\bar H}(x,y,z)$.

    One obvious source for the difference between $q_f(x)$ and $\bar
    q_f(x)$ is the valence quark contribution. Although this influences
    only $u$ and $d$, it makes the ratio of the contributions from $u$,
    $d$ and $s$ to $H$ different from the corresponding ratio for the
    contributions from $\bar u$, $\bar d$ and $\bar s$ to $\bar H$.
    As we will see clearly from Fig.~\ref{fig:SfH} in next subsection, $S_f^H$ is very much different for
    different $f$, and such a different ratio leads to different $P_H$ and $P_{\bar H}$.

    Clearly, valence quark contributions are negligible at very small $x$.
    We therefore expect that its influence becomes negligible at
    very high energies. Also, since the influence is determined by the ratio of $u(x)$,
    $d(x)$ and $s(x)$, the results can be quite sensitive to the
    forms of the parton distributions.

    The unpolarized parton distribution functions $q_f(x)$ are
    determined from unpolarized deep inelastic scattering and other
    related data from unpolarized experiments. There are different sets
    available in the parton distribution function library (PDFLIB~\cite{pdflib}) package.
    Although the qualitative features are all the same, there are
    differences in the fine structure, which
    may influence the difference between $R_f^H(x,y,z)$ and
    $R_{\bar f}^{\bar H}(x,y,z)$ and lead to different results in
    $P_H$ and $P_{\bar H}$. We will study this in next sections.

    Another source of the difference between $q_f(x)$ and $\bar q_f(x)$ is
    the asymmetry in nucleon sea and anti-sea distributions. Physical
    picture for such an asymmetry was proposed~\cite{Signal:1987gz,Brodsky:1996hc,
    Burkardt:1991di,Holtmann:1996be,Christiansen:1998dz,Cao:1999da} and models or
    parameterizations exist~\cite{Olness:2003wz,Lai:2007dq}. 
    This should be the dominant source for the
    difference between $q_f(x)$ and $\bar q_f(x)$ at very small $x$ and
    can be better studied at higher energies if it indeed leads to a
    significant difference between $R_f^H(x,y,z)$ and
    $R_{\bar f}^{\bar H}(x,y,z)$ thus a significant difference between
    $P_H(z)$ and $P_{\bar H}(z)$.
    We will also make calculations for this case in Sec. III. 

    \subsection{Spin transfer in fragmentation process}

    Spin transfer in fragmentation is defined in Eq.~(\ref{eq:Stransfer})
    and is given by the polarized fragmentation functions $\Delta
    D^H_f(z)$. For explicitly, we only consider the fragmentation
    process $q_f\to H+X$. We should note that, when writing the
    factorization theorem in the way as given in
    Eq.~(\ref{eq:factorization}), we assume that the fragmentation
    function is defined inclusively for the fragmentation process
    $q_f\to H+X$. It should include all the contributions from all the
    decay processes including strong as well as other decay processes.
    However, to study the physics behind it, it is useful to divide it
    into the directly produced part and the decay contributions. It has
    been widely used in studying the fragmentation functions in
    unpolarized processes and has been outlined in different
    publications~\cite{Liu:2001yt,Liu:2000fi,Xu:2002hz,Zuotang:2002ub,
    Dong:2005ea,Xu:2004es,Xu:2005ru,Chen:2007tm}. 
    Here, for completeness, we summarize the
    major equations in the following. 
    
    According to this classification,  we write,
    \begin{equation}
    D^H_f(z)=D^H_f(z;dir)+D^H_f(z;dec),
    \end{equation}
    where the $D^H_f(z;dir)$ and $D^H_f(z;dec)$ are the directly produced and decay contribution part respectively.
    The decay contribution can be calculated by,
    \begin{equation}
    D^H_f(z;dec)=\sum_j\int dz'K_{H,H_j}(z,z')D^{H_j}_f(z')
    \label{eq:DfHdec}
    \end{equation}
    where the kernel function $K_{H,H_j}(z,z')$ is the probability for $H_j$ with the
    fractional momentum $z'$ to decay into a $H$ with fractional momentum $z$,
    and, e.g, for a two body decay $H_j\to H+M$, it is given by,
    \begin{equation}
    K_{H,H_j}(z,z')=\frac{N}{E_j}{\rm Br}(H_j\to H+M)\delta(p.p_j-m_jE^*),
    \end{equation}
    where ${\rm Br}(H_j\to H+M)$ is the decay branching ratio, $N$ is the normalization constant,
    and $E^*$ is the energy of $H$ in the rest frame of $H_j$, and $m_j$ is the mass of $H_j$.

    Similarly, in the polarized case, we have,
    \begin{equation}
    \Delta D^H_f(z)=\Delta D^H_f(z;dir)+\Delta D^H_f(z;dec),
    \label{eq:deltaDfH2}
    \end{equation}
    and the decay part is given by,
    \begin{equation}
    \Delta D^H_f(z;dec)=\sum_j \int dz't_{H,H_j}^DK_{H,H_j}(z,z')\Delta D^{H_j}_f(z')
    \label{eq:deltaDfHdec}
    \end{equation}
    where $t_{H,H_j}^D$ is a constant called the decay spin transfer which is independent of the $H_j$ produced process,
    and is e.g. discussed and given in Table 2 of Ref.~\cite{Liu:2000fi}.

    We should note that, in Eqs.~(\ref{eq:DfHdec}) and~(\ref{eq:deltaDfHdec}), when calculating different decay contributions, 
    we have added the contributions from different hyperon decays incoherently. 
    This is what one often does in calculating inclusive quantities where the interferences are usually small 
    because of the small contributions from different channels to exactly the same final state at exactly 
    the same phase space points. 

    \subsubsection{Modeling $\Delta D_f^H(z,dir)$}

    Since fragmentation is a non-perturbative process, the fragmentation
    function can not be calculated using perturbative QCD. 
    At present, we have to invoke parameterization and/or phenomenological models. 
    There are already data available~\cite{Buskulic:1996vb,Ackerstaff:1997nh,Astier:2000ax,
    Adams:1999px,Airapetian:1999sh,Alexakhin:2005dz,Sapozhnikov:2005sc,Xu:2005js} 
    that can be used to extract information on
    the polarized fragmentation functions $\Delta D_f^H(z,dir)$ but
    still far away from giving a good control of the form of it. 
    At this stage, phenomenological models are quite useful in particular in
    obtaining some guide for experiments. In this connection, the model
    invoking calculation method according to the origins of hyperon is
    very practical and successful~\cite{Gustafson:1992iq,Boros:1998kc,Liu:2001yt,Liu:2000fi,
    Xu:2002hz,Zuotang:2002ub,Dong:2005ea,Xu:2004es,Xu:2005ru,Chen:2007tm} . 
    In this model, one classifies the
    directly produced hyperons into the following two categories: (A)
    those which contain the initial quark $q_f$ and (B) those which do
    not contain the initial quark, i.e.,
    \begin{equation}
    D_f^{H}(z;dir)=D_f^{H(A)}(z)+D_f^{H(B)}(z),
    \end{equation}
    \begin{equation}
    \Delta D_f^{H}(z;dir)=\Delta D_f^{H(A)}(z)+\Delta D_f^{H(B)}(z).
    \end{equation}
    It is assumed that those do not contain the initial quark are unpolarized, so that,
    \begin{equation}
    \Delta D_f^{H(B)}(z)=0.
    \label{eq:DfB}
    \end{equation}
    The polarization then originates only from category (A) and is given by,
    \begin{equation}
    \Delta D_f^{H(A)}(z)=t^F_{H,f} D_f^{H(A)}(z),
    \label{eq:DfA}
    \end{equation}
    in which $t^F_{H,f}$ is known as the fragmentation spin transfer factor and is taken as a constant given by,
    \begin{equation}
    t_{H,f}^F=\Delta Q_f/n_f,
    \label{eq:tHf}
    \end{equation}
    where $\Delta Q_f$ is the fractional spin contribution of a quark with flavor $f$ to the spin of the hyperon,
    and $n_f$ is the number of valence quarks of flavor $f$ in $H$.

    The model is very practical and useful for the following reason: In
    the recursive cascade hadronization models, such as Feynman-Field
    type fragmentation models~\cite{Field:1977fa} where a simple elementary
    process takes place recursively, $D_f^{H(A)}(z)$ and $D_f^{H(B)}(z)$
    are well defined and determined. In such models, $D_f^{H(A)}(z)$ is
    the probability to produce a first rank $H$  which is usually
    denoted by $f_{q_f}^H(z)$ and is well determined by unpolarized
    reaction data. Hence, the $z$-dependence $\Delta D$ given above is
    obtained completely from the unpolarized fragmentation functions,
    which are empirically known. The only unknown is the spin transfer
    constant $t^F_{H,f}=\Delta Q_f/n_f$.
    By using either the SU(6) wave function or polarized deep-inelastic lepton-nucleon scattering
    data, one obtains two distinct expectations $\Delta Q_f$,
    the so-called SU(6) and DIS expectations, see table 1 of~\cite{Liu:2000fi}.

    This approach has been applied to different hyperons/anti-hyperons
    in different reactions such as $e^+e^-$,  SIDIS and $pp$
    collisions~\cite{Gustafson:1992iq,Boros:1998kc,Liu:2001yt,Liu:2000fi,Xu:2002hz,Zuotang:2002ub,
    Dong:2005ea,Xu:2004es,Xu:2005ru,Chen:2007tm} and
    compared with data~\cite{Buskulic:1996vb,Ackerstaff:1997nh,Astier:2000ax,
    Adams:1999px,Airapetian:1999sh,Alexakhin:2005dz,Sapozhnikov:2005sc,Xu:2005js}.
    The current experimental accuracy does not allow one to distinguish
    between the expectations for $t^F_{H,f}$ based on the SU(6) and
    DIS pictures. However, the $z$-dependence of the available data on
    $\Lambda$ polarization is well described~\cite{Liu:2000fi}.

    \subsubsection{Numerical results for $S_f^H(z)$}
\begin{figure} [htb]
    \resizebox{0.8\textwidth}{!}{\includegraphics{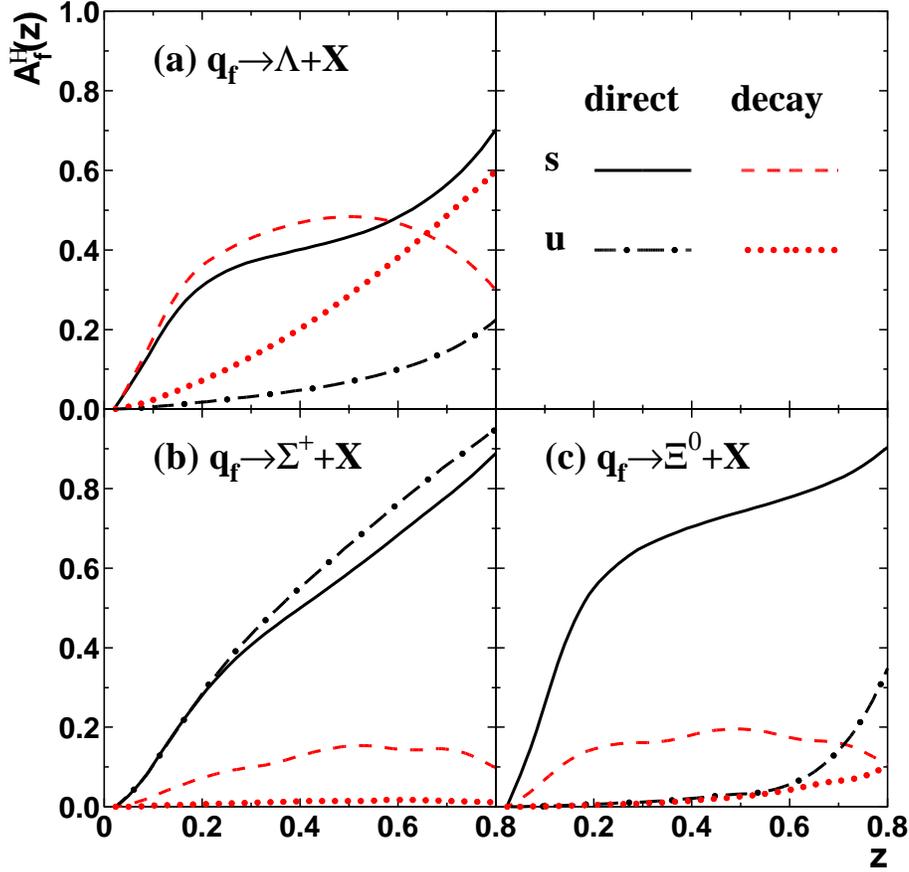}}
    \caption{(color online) First rank contributions $A_f^H(z,dir)$ and $A_f^H(z,dec)$ in quark fragmentation 
    to the productions of different hyperons as the functions of $z$. 
    The results are extracted from $e^+e^-$ process with $\sqrt{s}$=200 GeV using {\sc pythia}.  }
    \label{fig:AfH}       
    \end{figure}

    Using the definition given in Eq.~(\ref{eq:Stransfer}) and~(\ref{eq:deltaDfH2}), 
    we obtain the spin transfer for $q_f\to H+X$ as,
    \begin{equation}
    S_f^H(z)=S_f^H(z,dir)+S_f^H(z,dec).
    \end{equation}
    In the model described above, we have,
    \begin{equation}
    S_f^H(z,dir)=t_{H,f}^FA_f^H(z,dir),
    \end{equation}
    \begin{equation}
    A_f^H(z,dir)=f_{q_f}^H(z)/D^H_f(z),
    \end{equation}
    If we do not consider successive decay, we have,
    \begin{equation}
    S_f^H(z,dec)=\sum_jt_{H_j,f}^Ft^D_{H,H_j}A^H_{f,H_j}(z,dec),
    \end{equation}
    \begin{equation}
    A^H_{f,H_j}(z,dec)=\int dz'K_{H,H_j}(z,z')f_{q_f}^{H_j}(z')/D^H_f(z).
    \end{equation}
    We call $A^H_f(z,dir)$ and $A^H_{f,H_j}(z,dec)$ first rank contributions and note that 
    both of them are determined by the unpolarized fragmentation functions. 
    As an example, we calculated them using Lund fragmentation model\cite{Andersson:1983ia}
    as implemented in the Monte-Carlo event generator {\sc pythia}~\cite{Sjostrand:1993yb}. 
    The results are given in Fig.~\ref{fig:AfH}. 
    Because isospin symmetry is valid here, so we have relations such as 
    $A^\Lambda_u(z,dir)=A^\Lambda_d(z,dir)$, 
    $A^{\Sigma^+}_u(z,dir)=A^{\Sigma^-}_d(z,dir)$, 
    $A^{\Xi^0}_u(z,dir)=A^{\Xi^-}_d(z,dir)$, etc.  
    Hence, we only show the $u$ and $s$ contributions to $\Lambda$, $\Sigma^+$ and $\Xi^0$. 
    All the others from $u$, $d$ and $s$-quark to the $J^P=(1/2)^+$ hyperons can be obtained using 
    such relations. 
    We see, first of all,  that the decay contributions to $\Lambda$ are large 
    but those to $\Sigma$ and $\Xi$ are negligible. 
    We also see that $s$-quark contributions are large in general because those from $u$ or $d$ 
    have strangeness suppression, a well known factor in fragmentation process. 
    
\begin{figure}[htb] 
    \resizebox{0.8\textwidth}{!}{\includegraphics{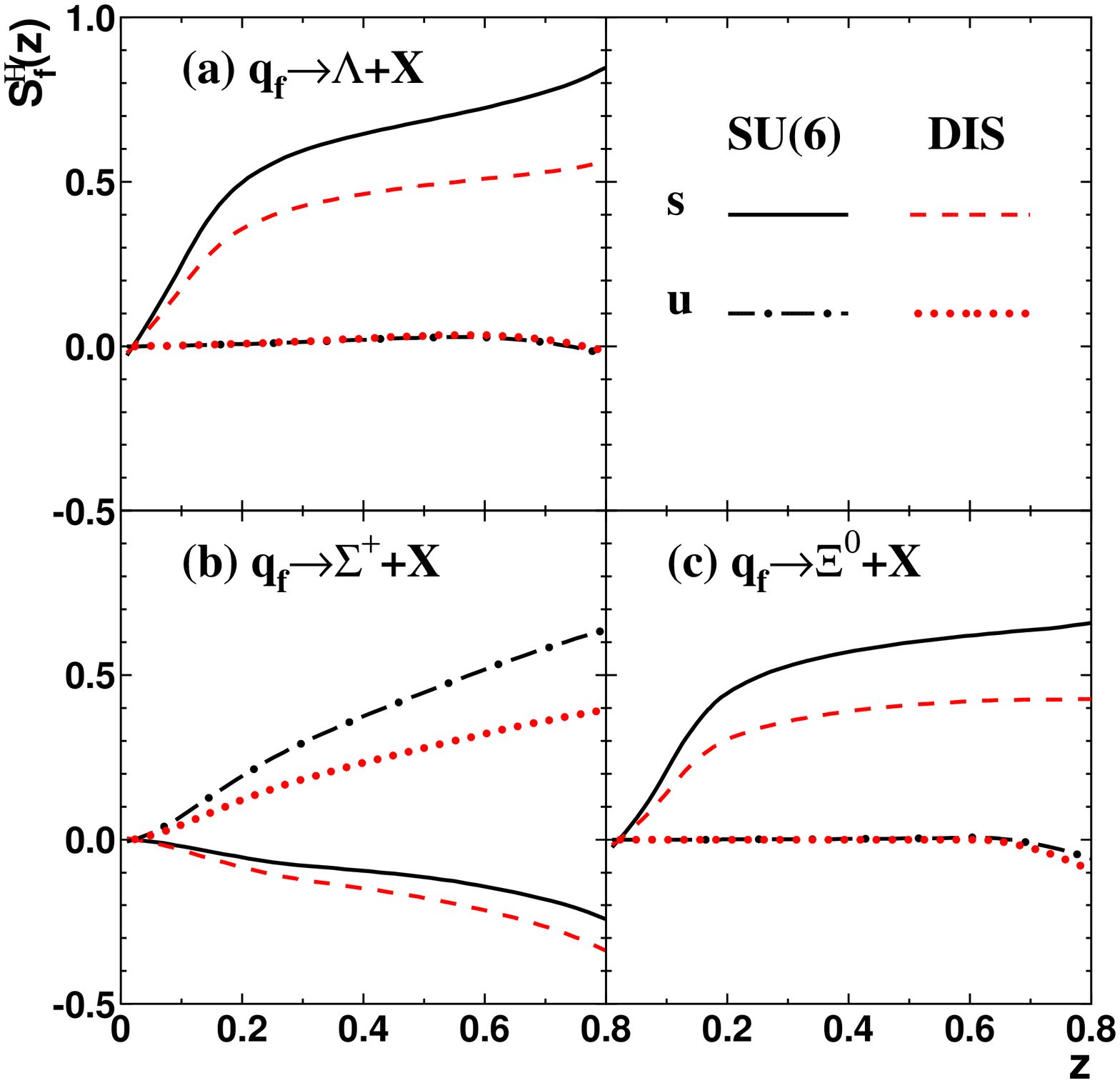}}
    \caption{(color online) Spin transfer $S_f^H(z)$ in the fragmentation process $q_f\to H+X$. }
    \label{fig:SfH}       
    \end{figure}

     Multiplying by the corresponding spin transfer factors $t^F_{H,f}$ and $t^D_{H,H_j}$, 
    we obtain the corresponding $S_f^H(z)$ as shown in Fig.~\ref{fig:SfH}. 
    
     We see that,  for different flavors,  $S_f^H(z)$ differs very much from each other 
     not only because of the difference in the first rank contributions as shown in  Fig.~\ref{fig:AfH}, 
     but also because of the differences in the spin transfer factors $t_{H,f}^F$ and $t_{H,H_j}^D$. 
     As an example, we see $S_u^{\Sigma^+}$ is positive and large while $S_s^{\Sigma^+}$ 
     is negative and the magnitude is smaller than  $S_u^{\Sigma^+}$.
    We also note that isospin symmetry is valid here so that we have a series of relations such as,
    $S_u^\Lambda(z)=S_d^\Lambda(z)$,
    $S_u^{\Sigma^+}(z)=S_d^{\Sigma^-}(z)$,
    $S_u^{\Xi^0}(z)=S_d^{\Xi^-}(z)$,
    $S_s^{\Sigma^+}(z)=S_s^{\Sigma^-}(z)$,
    and  $S_s^{\Xi^0}(z)=S_s^{\Xi^-}(z)$.

     \subsubsection{Comparing $S_f^H(z)$ with $S_{\bar f}^{\bar H}(z)$}
    For the fragmentation function, the directly produced part is controlled by strong interaction
    where charge conjugation symmetry is valid. Hence, we have,
    \begin{equation}
    D_f^H(z,dir)=D_{\bar f}^{\bar H}(z,dir),
    \end{equation}
    \begin{equation}
    \Delta D_f^H(z,dir)=\Delta D_{\bar f}^{\bar H}(z,dir),
    \end{equation}
    For the decay contributions, we have processes controlled by strong
    or electromagnetic interactions. In these processes, we still have charge conjugation symmetry
    so that similar equations as given above are valid.
    However, there are also weak decay processes that play a role.
    In a weak process, charge conjugation symmetry may be violated.
    There are a few weak decay processes that we need to take into account and
    we can check them one by one.
    Fortunately, for all the weak decay processes that give significant contribution to
    the production of hyperons (anti-hyperons) in our interest, no significant violation
    in charge conjugation symmetry has been observed.
    We therefore neglect the influence and have approximately that,
    \begin{equation}
    D_f^H(z)=D_{\bar f}^{\bar H}(z),
    \end{equation}
    \begin{equation}
    \Delta D_f^H(z)=\Delta D_{\bar f}^{\bar H}(z),
    \end{equation}
    We thus also have,
    \begin{equation}
    S_f^H(z)=S_{\bar f}^{\bar H}(z),
    \end{equation}
    \begin{equation}
    S_{\bar f}^H(z)=S_f^{\bar H}(z),
    \end{equation}

    We see, under such circumstances, we expect no significant
    difference between the spin transfer in quark fragmentation and that in
    anti-quark fragmentation. A significant difference between final
    hyperon and anti-hyperon polarization can only be from the difference
    between $P_f(x,y)$ and $P_{\bar f}(x,y)$ and/or that between
    $R_f^H(x,y,z)$ and $R_{\bar f}^{\bar H}(x,y,z)$. We recall that, in
    the case of $P_T=0$,  $P_f(x,y|P_T=0)=P_{\bar f}(x,y|P_T=0)=P_bD_L(y)$,
    the only source for such a difference is the difference between
    $R_f^H(x,y,z)$ and $R_{\bar f}^{\bar H}(x,y,z)$, which we discussed in Sec. IIC.

    As an example, we see that, in the model described in Sec.IID(2), 
    \begin{equation}
    S_{\bar f}^H(z)=S_f^{\bar H}(z)=0.
    \end{equation}
    \begin{equation}
    S_f^H(z)=S_{\bar f}^{\bar H}(z)=t_{H,f}^FA_f^H(z,dir)+\sum_jt_{H_j,f}^Ft_{H,H_j}^DA^H_{f,H_j}(z,dec).
    \end{equation}
    Charge conjugation symmetry is indeed valid here and the numerical results are given in Fig.~\ref{fig:SfH}. 
    We emphasize here that, as can be seen from Fig.~\ref{fig:SfH},
     for different flavor $f$, $S_f^H(z)$ differs very much from each other. 
     This makes the value of the polarization of final hyperon
    sensitive to  $R_f^H(x,y,z)$. Hence, measuring $P_H$ is a good way
    to study the fine behavior of $R_f^H(x,y,z)$.

    \subsection{A practical way of the calculations}
    As shown by Eqs.~(\ref{eq:polHsimp}-\ref{eq:RfHpol}),
    the calculations of $P_H$ and/or $P_{\bar H}$
    involve the contributions from different flavor $f$ and $\bar f$,
    each of them is a convolution of quark distributions, polarization
    fragmentation function and other kinematic factors originating from
    the $eq$ scattering etc.
    Using the parton distributions from PDFLIB~\cite{pdflib},
    the perturbative calculation results for the differential cross section for
    $eq\to eq$, and the parameterization for the fragmentation functions,
    we can in principle calculate the contribution in a
    straight-forward manner.
    However, in view of the number of different flavor $f$ and $\bar f$ involved,
    all the different decay contributions and the difficulties and/or uncertainties
    in obtaining the fragmentation functions, the calculations are almost impossible
    without radical approximations.
    On the other hand, all these information for unpolarized reactions are implemented in
    the Monte-Carlo event generators such as {\sc lepto}~\cite{Ingelman:1996mq} so
    that the corresponding unpolarized cross section can be calculated
    conveniently using such Monte-Carlo programs.
    Such Monte-Carlo event generators have been developed since 1980s
    and have been tested by enormous amount of unpolarized experiments
    and the parameters in the models have been adjusted to fit all the data.
    They provide a useful tool to make predictions for out-coming experiments
    and are widely used in the community.
    The Monte-Carlo event generators for
    unpolarized high energy reactions have also been used to calculate
    the corresponding unpolarized parts in calculating the polarizations
    of the produced hadrons in 
    literature~\cite{Gustafson:1992iq, Boros:1998kc,
    Liu:2001yt,Liu:2000fi,Xu:2002hz,Zuotang:2002ub,Dong:2005ea,Xu:2004es,Xu:2005ru,Chen:2007tm,
    Ellis:1995fc,Kotzinian:1997vd,Ellis:2002zv,Ellis:2007ig}.
    
    To show how this is carried out, we take the polarization of $H$ in
    the case of $P_T=0$ as an example and re-write Eq.~(\ref{eq:polH}) as,
    \begin{equation}
    P_{H}(z|P_T=0)=\frac{\sum_{f,\alpha}e^2_f\int dxdy  P_bD_L(y) K(x,y)
    q_{f}(x) D_f^{H(\alpha)}(z) S_f^{H(\alpha)}}
    {\sum_fe^2_f\int dxdy K(x,y) \left[q_f(x)D^H_f(z)+\bar q_f(x)D^H_{\bar f}(z) \right] },
    \label{eq:polHMC}
    \end{equation}
    where $\alpha=A$ through $D$ denoting the different origins of $H$:
    (A) directly produced and contain $q_f$; (B) directly produced but
    do not contain $q_f$; (C) decay product of $H_j$ which is directly
    produced and contain $q_f$; (D)  decay product of $H_j$ which is
    directly produced and do not contain $q_f$. $S_f^{H(\alpha)}=\Delta
    D_f^{H(\alpha)}(z)/D_f^{H(\alpha)}(z)$ is the spin transfer factor
    in fragmentation for each contribution. We see from
    Eqs.~(\ref{eq:deltaDfHdec}), (\ref{eq:DfA}) and (\ref{eq:DfHdec}) that,
    $S_f^{H(A)}=t^F_{H,f}$, $S_f^{H(C)}=t^F_{H_j,f}t^D_{H,H_j}$ and
    $S_f^{H(B)}=S_f^{H(D)}=0$.
    Denote the relative contribution from origin ($\alpha$) by,
    \begin{equation}
    R_f^{H(\alpha)}(x,y,z)=\frac{e^2_f K(x,y) q_f(x) D_f^{H(\alpha)}(z) }
    {\sum_fe^2_f\int dxdy K(x,y) \left[q_f(x)D^H_f(z)+\bar q_f(x)D^H_{\bar f}(z) \right] },
    \label{eq:Ralpha}
    \end{equation}
    and we have, 
    \begin{equation}
    P_{H}(z|P_T=0)=\sum_{f,\alpha}\int dxdy  R_f^{H(\alpha)}(x,y,z) P_bD_L(y)  S_f^{H(\alpha)}.
    \label{eq:polHMC2}
    \end{equation}
    We see that the relative production weight $R_f^{H(\alpha)}(x,y,z)$ defined in 
    Eq.~(\ref{eq:Ralpha}) is independent of the polarization and 
    can be calculated using a Monte-Carlo event generator.
    In the right hand side of Eq.(\ref{eq:polHMC2}), the quark polarization $P_bD_L(y)$ and
    the spin transfer constant $S_f^{H(\alpha)}$ are two known quantities and they are the places
    where information on polarization comes in.
    In practice, we generate a $ep$ collision event using an event generator.
    We study the final state hadrons and search for the hyperon $H$ under study.
    After we find a $H$ in the considered kinematic region,
    we  calculate $D_L(y)$ and $S_f^{H(\alpha)}$ by tracing back the origin of
    the $H$ using information recorded in the Monte-Carlo program.

    Usually a Monte-Carlo event generator is tested by the existing data at different energies,
    and is expected that it can give a reasonable description of the unpolarized quantities at a given energy
    provided that the physics does not have sudden changes at that energy.
    We therefore use such Monte-Carlo event generator for such analysis in the following.
    Such a calculation method is not only the most convenient way 
    available currently of calculating the fragmentation functions
    and the contributions from different hard scattering processes but also
    the most convenient way to include the contributions from all different decay processes.

    \subsection{A lower energy effect}
    At lower energies, there is another effect that relates to the valence quark contribution
    and may cause a difference between $P_H$ and $P_{\bar H}$ in
    $e^-+N\to e^-+H({\rm or}\ \bar H)+X$, i.e.,
    the contribution from the hadronization of the remnant of target nucleon.
    It has been pointed out first in~\cite{Zuotang:2002ub} that contribution of
    the hadronization of target remnant is important to hyperon
    production even for reasonably large $x_F$ at lower energies. 
    (Here, $x_F$ is the Feynman-$x$ in the c.m. frame of the $\gamma^*p$ system, 
    which is approximately equal to $z$ at high energy and small $p_T$.)
    The effect has been confirmed by the calculations presented in~\cite{Ellis:2002zv}.
    It has been shown that~\cite{Zuotang:2002ub} at the CERN NOMAD energies,
    contributions from the hadronization of nucleon
    target remnant dominates hyperon production at $x_F$ around zero.
    It is impossible to separate the contribution of the struck quark
    fragmentation from those of the target remnant fragmentation.
    In this case, the factorization theorem given in Eq.~(\ref{eq:factorization}) is broken down
    and the concept of independent fragmentation is no more valid.
    Clearly, this effect can be different for hyperon and anti-hyperon
    production since the target remnant contribution comes mainly from
    the fragmentation of the valence di-quark.
    It contributes quite differently for hyperons and anti-hyperons.
    This can have a large influence at lower energies, but the influence
    should vanish at high energies where current fragmentation can well
    be defined.
    As have already demonstrated in Ref.~\cite{Dong:2005ea}, this low energy effect has already 
    little influence on the results at COMPASS energy, in particular when studying 
    the difference between $P_H$ and $P_{\bar H}$. 
    We therefore do not consider this effect in the following of this paper.

    \section{$P_H$ and $P_{\bar H}$ in SIDIS with polarized beam and unpolarized target}
    
    By using the method presented in last section, we can calculate the
    polarizations of hyperon and anti-hyperon in SIDIS with the aid of a Monte-Carlo event generator. 
    We use the latest version of {\sc lepto 6.5.1}~\cite{Ingelman:1996mq} 
    based on the Lund string fragmentation model~\cite{Andersson:1983ia} in the following. 
    We study  SIDIS with longitudinally polarized 
    electron (muon) beam and unpolarized target in this section and make calculations of 
    $P_H$ and $P_{\bar H}$ in the case that a symmetric strange sea $s(x)$ and anti-sea distribution $\bar s(x)$ is assumed 
    and in the case that an asymmetry  between $s(x)$ and $\bar s(x)$ is taken into account separately.
    
    \subsection{Results with symmetric strange sea and anti-sea distributions}

    Our calculations in this case are made by using a parameterization of PDF's (parton distribution functions) 
    obtained in PDFLIB where no asymmetry between strange sea and anti-sea distribution is taken into account.
    It is obvious that different sets of parameterizations of PDF's may 
    have influence on our results of $P_H$ and $P_{\bar H}$.
    In the following, we first present the results obtained using CTEQ2L and study 
    the influence of different sets of parameterizations later in this section.
    
    \subsubsection{Results at COMPASS energy}
\begin{figure}[htb] 
    \resizebox{0.8\textwidth}{!}{\includegraphics{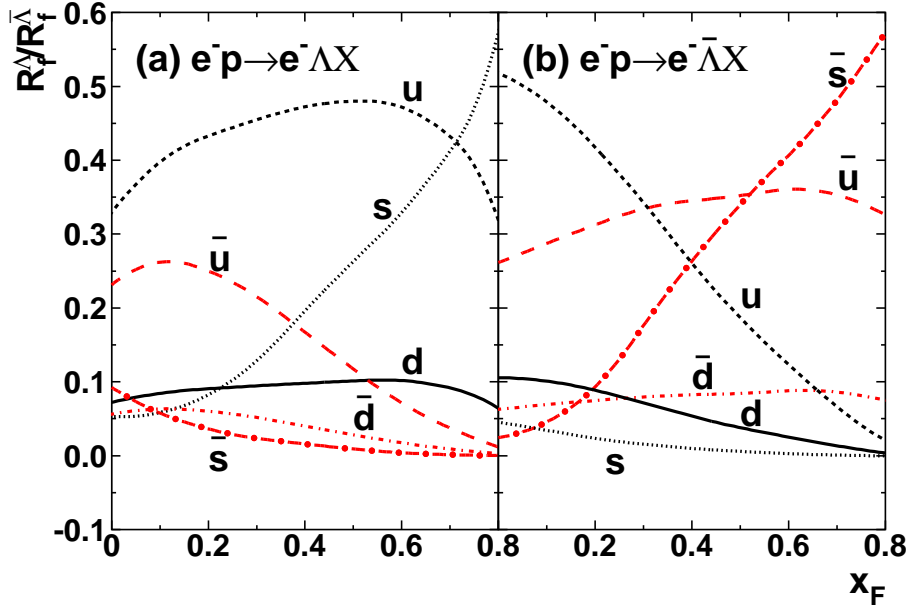}}
    \caption{(color online) Relative weights $R_f^{\Lambda}$ and $R_{f}^{\bar{\Lambda}}$  
     as functions of $x_{F}$ at COMPASS energy $\sqrt{s}$=17.35 GeV 
     for different flavors $f=u, d, s, \bar u, \bar d,$ and $\bar s$ respectively.}
    \label{fig:lamborg}       
    \end{figure}

    To compare with the available experimental data, we first make calculations  
    in the same kinematic region as in the COMPASS experiment~\cite{Alexakhin:2005dz,Sapozhnikov:2005sc}, i.e., 
    $Q^2>1$GeV$^{2}$, and $0.2<y<0.9$ with $\mu$ beam energy of 160 GeV
    and beam polarization $P_{\mu}=-0.76$.
     As we have mentioned in last section,  when the target is
    unpolarized, the difference between the polarizations of $H$ and
    $\bar{H}$ comes only from the relative weight $R_f^{H}$ and
    $R_{\bar{f}}^{\bar{H}}$ as given by Eqs.(\ref{eq:polHsimp})
    and (\ref{eq:polHbarsimp}).
    We therefore first calculated $R_f^{\Lambda}$ and $R_{f}^{\bar{\Lambda}}$ 
    and show the results obtained in Figs.~\ref{fig:lamborg}(a) and (b) respectively.
    
    From Figs.~\ref{fig:lamborg}(a) and (b), we see the following features: 
    
    (1) With increasing $x_F$, $R_s^\Lambda$ and $R_{\bar s}^{\bar\Lambda}$ increase fast, 
    $R_{u,d}^\Lambda$ or $R_{\bar u,\bar d}^{\bar\Lambda}$ vary slowly, 
     while $R_{\bar q}^\Lambda(x_F)$ and $R_q^{\bar\Lambda}(x_F)$ decrease, 
     in particular in the large $x_F$ region. 
     This is because $R_{u,d,s}^\Lambda$ and $R_{\bar u,\bar d,\bar s}^{\bar\Lambda}$  
     have the first rank contributions while $R_{u,d,s}^{\bar\Lambda}$ and $R_{\bar u,\bar d,\bar s}^\Lambda$ 
     do not and the first rank contributions to $R_{u,d}^\Lambda$ and $R_{\bar u,\bar d}^{\bar\Lambda}$ 
     have strangeness suppression compared to $R_{s}^\Lambda$ and $R_{\bar s}^{\bar\Lambda}$ . 
    
    (2) The shape of $R_u^\Lambda(x_F)$ is similar to $R_d^\Lambda(x_F)$ but the former is in general 
    much larger than the latter. This is not only because of the larger contribution from the valence to $u$ 
    but also because of that the electric charge squared factor for $u$ is 4 times as large as that for $d$. 
    Similar relations hold for $\bar u$ and $\bar d$ contributions and also 
    for those to $\bar\Lambda$. 
    
    (3) There is in general quite a large difference between 
    $R_f^\Lambda$ and $R_{\bar f}^{\bar\Lambda}$ for a given $f$, i.e., 
    the charge conjugation symmetry is not hold here. 
    The difference is particularly large for $R_u^\Lambda$ and $R_{\bar u}^{\bar\Lambda}$. 
    This is a characteristics of the contribution from the valence quark. 
    
\begin{figure}[htb]
    \resizebox{0.5\textwidth}{!}{\includegraphics{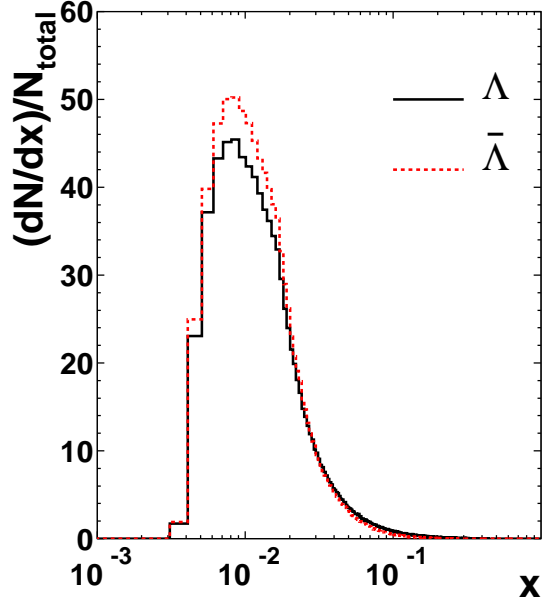}}
    \caption{(color online) The $x$-distribution of the struck quark or anti-quark that leads to the production of 
    $\Lambda$ or $\bar{\Lambda}$ in the kinematic region of $x_F>0$ at COMPASS energy $\sqrt{s}$=17.35 GeV.}
    \label{fig:xdepcomp}       
    \end{figure}

    To understand that the valence quark contributions are still large at COMPASS energy, 
    Ref.~\cite{Dong:2005ea} has calculated the $x$-values 
    of the struck quark and/or anti-quark. We did similar calculations in exactly the COMPASS 
    kinematic region as described above and obtain the results shown in Fig.~\ref{fig:xdepcomp}. 
    We see that most of the $\Lambda$'s and $\bar{\Lambda}$'s are from the
    quark and anti-quark with momentum fraction $x$ around 0.01.
    In this $x$ region, the valence quark contributions are indeed still quite large 
    so that $u(x)>\bar{u}(x)$  and $d(x)>\bar{d}(x)$.
    This leads to much larger $u$ ($d$) contributions to $\Lambda$
    than the corresponding $\bar{u}$ ($\bar{d}$) contribution to $\bar\Lambda$.
    
\begin{figure}[htb]
    \resizebox{0.6\textwidth}{!}{\includegraphics{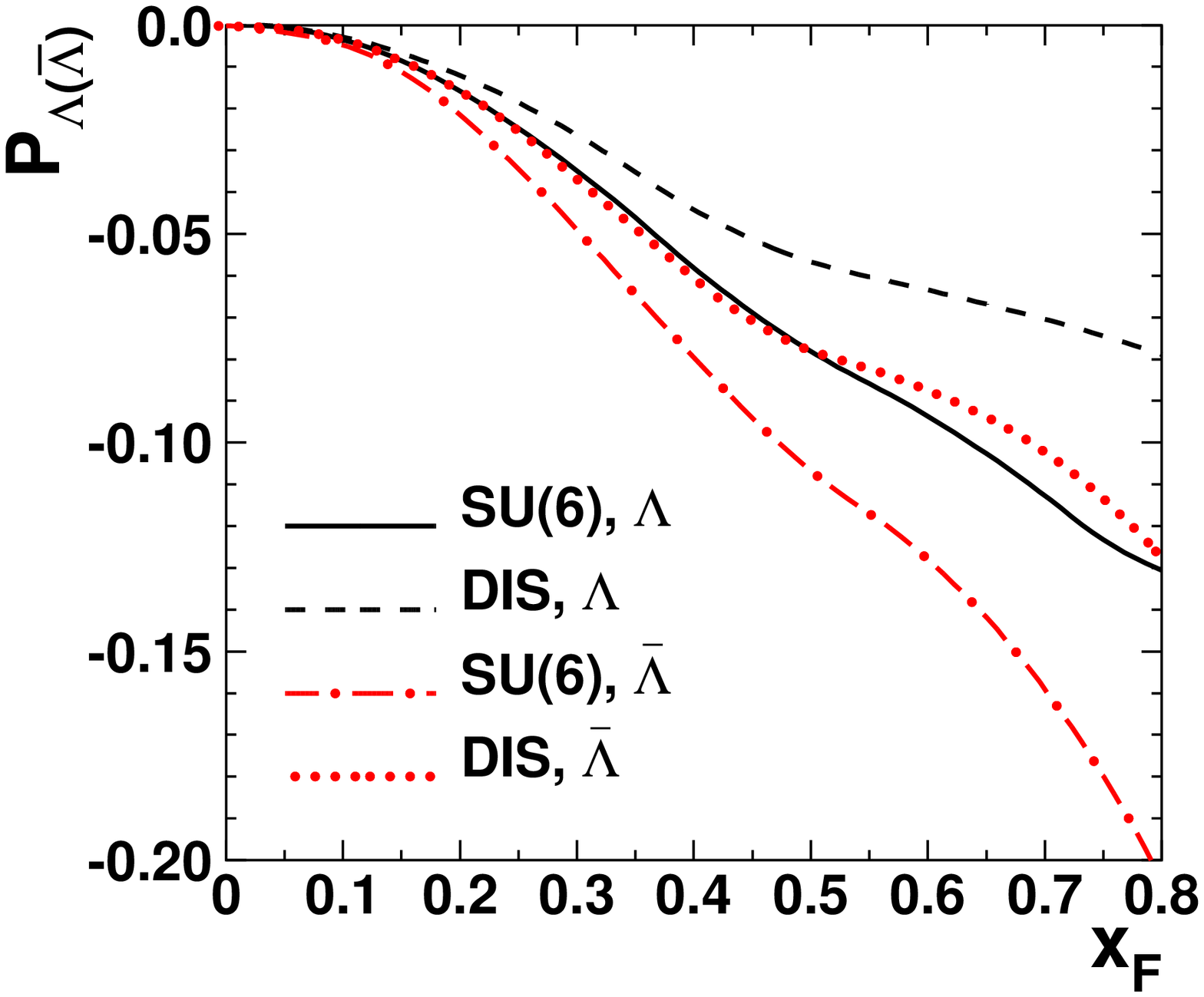}}
    \caption{(color online) Polarizations of $\Lambda$ and $\bar{\Lambda}$ as the
    functions of $x_{F}$ at COMPASS energy $\sqrt{s}$=17.35 GeV 
    obtained using symmetric strange sea and anti-sea distributions.}
    \label{fig:lambpolcompass}       
    \end{figure}

    We recall that the spin transfer in fragmentation $S_f^\Lambda=S_{\bar f}^{\bar\Lambda}$ 
    is quite different for $f=u$ or $d$ from that for $f=s$ (see Fig.\ref{fig:SfH}), 
    we thus expect that there is a significant 
    difference between $P_\Lambda$ and $P_{\bar\Lambda}$. 
    We calculate these polarizations using the spin transfer $S_f^\Lambda(z)=S_{\bar f}^{\bar\Lambda}(z)$ 
    described in last section, and show the results in Fig.~\ref{fig:lambpolcompass}. 
    We see that there are indeed some difference between  $P_\Lambda$ and $P_{\bar\Lambda}$ at the same $x_F$. 
    The magnitude of  $P_{\bar\Lambda}$ is larger than that of $P_\Lambda$. 
    This is because that the contribution from $u$-quark is larger and that $S_u^\Lambda$ is small and negative. 
      
    \begin{figure}[htb]
    \resizebox{0.6\textwidth}{!}{\includegraphics{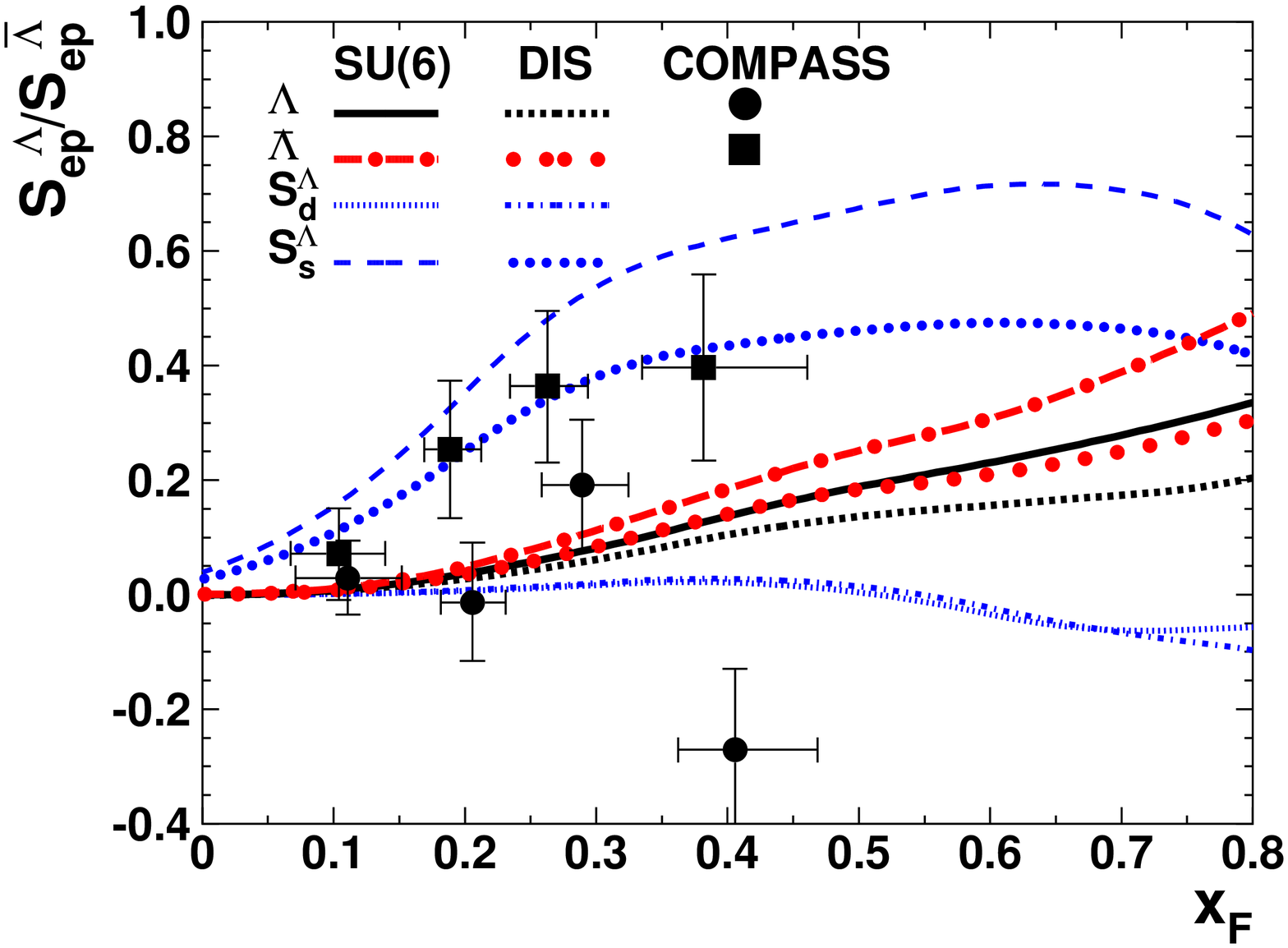}}
    \caption{(color online) Spin transfer $S_{ep}^{\Lambda / \bar{\Lambda}}$ as the function
    of $x_F$ at COMPASS energy $\sqrt{s}$=17.35 GeV obtained using  symmetric 
    strange sea and anti-sea distributions.
    The data points are taken from COMPASS [7].}
    \label{fig:polHterminalcompass}       
    \end{figure}
 
    To compare with the data from COMPASS\cite{Alexakhin:2005dz,Sapozhnikov:2005sc}, 
    we also calculate the spin transfer 
    $S_{ep}^\Lambda$ and $S_{ep}^{\bar\Lambda}$ in $e^-+p\to e^-+\Lambda (\bar\Lambda)+X$ 
    as given by Eq.~(\ref{eq:epspintransfer3}) and show the results in Fig.~\ref{fig:polHterminalcompass}. 
    We recall that the magnitude of the spin transfer of $s$ ($\bar{s}$) quark to $\Lambda$ ($\bar\Lambda$)
    in the fragmentation process is much larger than that of $u$ or $d$ ($\bar u$ or $\bar d$) 
    quark (see Fig.\ref{fig:SfH}),
    so $S_{ep}^\Lambda$ ($S_{ep}^{\bar \Lambda}$) takes its maximum
    when there is only contribution from strange quarks, i.e., $S_{ep}^\Lambda \rightarrow S_s^\Lambda$.
     In contrast,  $S_{ep}^{\Lambda}$ ($S_{ep}^{\bar \Lambda}$) reaches its
    minimum when there is only contribution from $u$ and/or $d$ ($\bar u$ and/or $\bar d$),  
    i.e., $S_{ep}^{\Lambda}\to S_{u}^{\Lambda}=S_{d}^{\Lambda}$. 
    These are the limits of $S_{ep}^\Lambda$ and  $S_{ep}^{\bar\Lambda}$ in $e^-+p\to e^-+\Lambda (\bar\Lambda)+X$.
    To show the range of $S_{ep}^\Lambda$ and  $S_{ep}^{\bar\Lambda}$ in the case that 
    the spin transfer model described in Sec.IID is used, we also show these two limits in the same figure.
    
    From Fig.~\ref{fig:polHterminalcompass}, 
    we see that, with a symmetric strange sea and anti-sea distribution, we still obtain some differences 
    between $S_{ep}^\Lambda$ and $S_{ep}^{\bar\Lambda}$ as functions of $x_F$. 
    But the differences seem not as large as those observed 
    by COMPASS collaboration\cite{Alexakhin:2005dz,Sapozhnikov:2005sc}. 
    From the limits $S_s^\Lambda$ and $S_u^\Lambda$, we see also that there are enough room 
    to fit the data by adjusting the relative weights of $s$ ($\bar s$) contributions 
    compared to those from $u$ and $d$ ($\bar u$ and $\bar d$). 
    
\begin{figure}[htb] 
    \resizebox{0.6\textwidth}{!}{\includegraphics{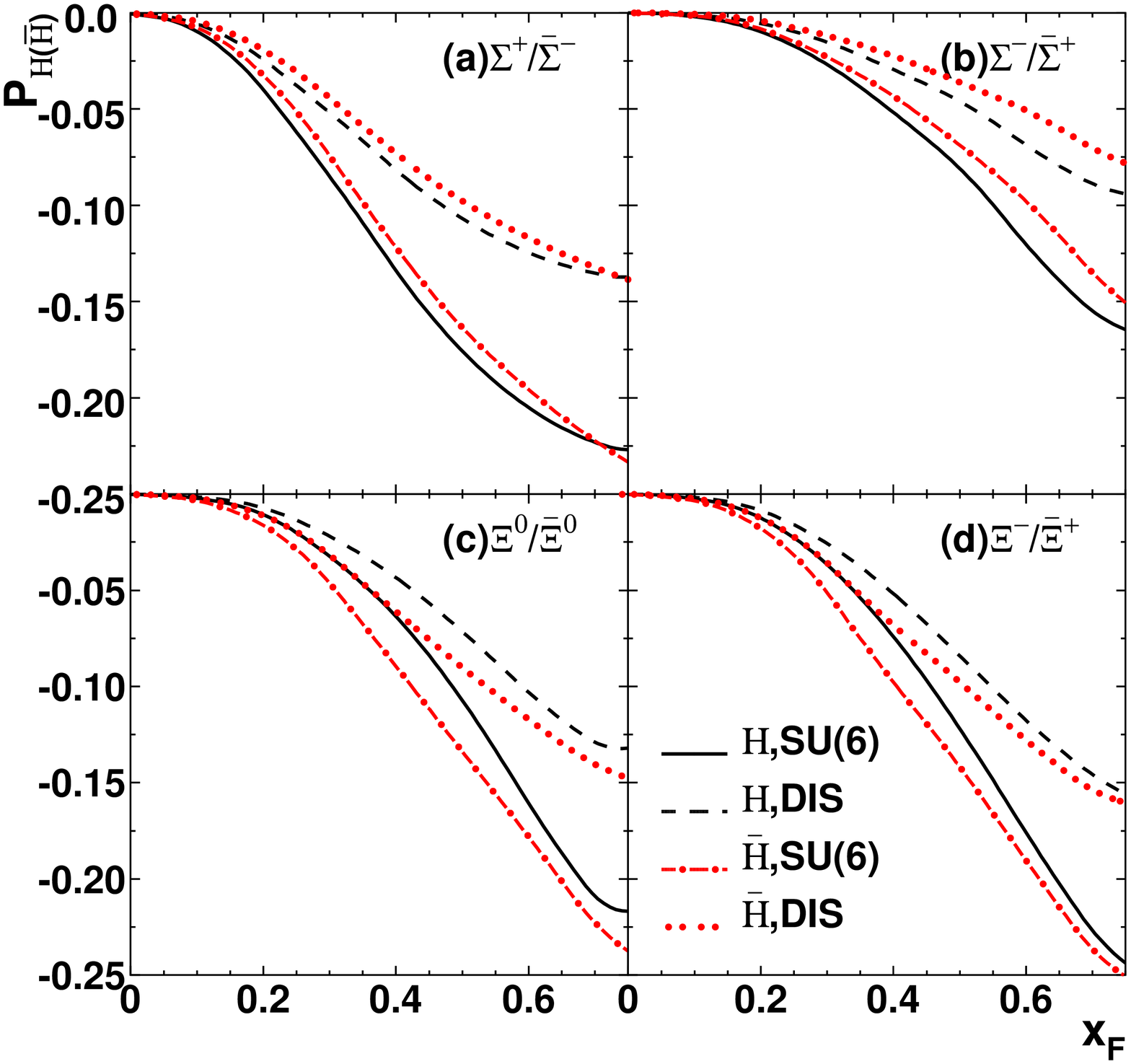}}
    \caption{(color online) Polarizations of hyperons and anti-hyperons as functions of $x_{F}$ at 
    COMPASS energy $\sqrt{s}$=17.35 GeV obtained using symmetric strange sea and anti-sea distributions. } 
    \label{fig:polhypcomp}       
    \end{figure}

    We have seen that the differences between $P_\Lambda$ and $P_{\bar\Lambda}$ in the case discussed in 
    this subsection come only from valence quark contributions. 
    The differences are due to that the relative contribution from 
    $s$ to $\Lambda$ is different from the relative contribution from $\bar s$ to $\bar\Lambda$. 
    If we extend the study to other $J^P=(1/2)^+$ hyperons such as $\Sigma^{\pm}$ and $\Xi$, 
    the situations can be different. 
    For example, for $\Sigma^+$ and its anti-particle $\bar\Sigma^-$, the production and 
    the polarization are dominated by $u$ and $\bar u$ contributions respectively. 
    Although valence quark contributions make $u$ dominance even stronger, 
    but the relative weights do not change much, even at COMPASS energy. 
    Similarly,  $\Sigma^-$ and $\bar\Sigma^+$ are dominated by $d$ and $\bar d$, 
   and $\Xi$ and $\bar\Xi$ are dominated by $s$ and $\bar s$ respectively. 
    We expect much smaller difference between $P_H$ and $P_{\bar H}$ for these hyperons.
    In Fig.~\ref{fig:polhypcomp}, we show the corresponding results at COMPASS energy.
    We see that the differences obtained between $P_H$ and $P_{\bar H}$ are indeed 
    much smaller than those for $\Lambda$ and $\bar{\Lambda}$.
    Since the decay contributions to these hyperons are almost negligible, 
    the calculations here are simpler and more clear. 
    This provides a rather clean test to see whether the difference between $P_\Lambda$ and $P_{\bar\Lambda}$ 
    are due to valence contributions. 
     
    \subsubsection{Results at eRHIC energy}
\begin{figure}[htb] 
    \resizebox{0.5\textwidth}{!}{\includegraphics{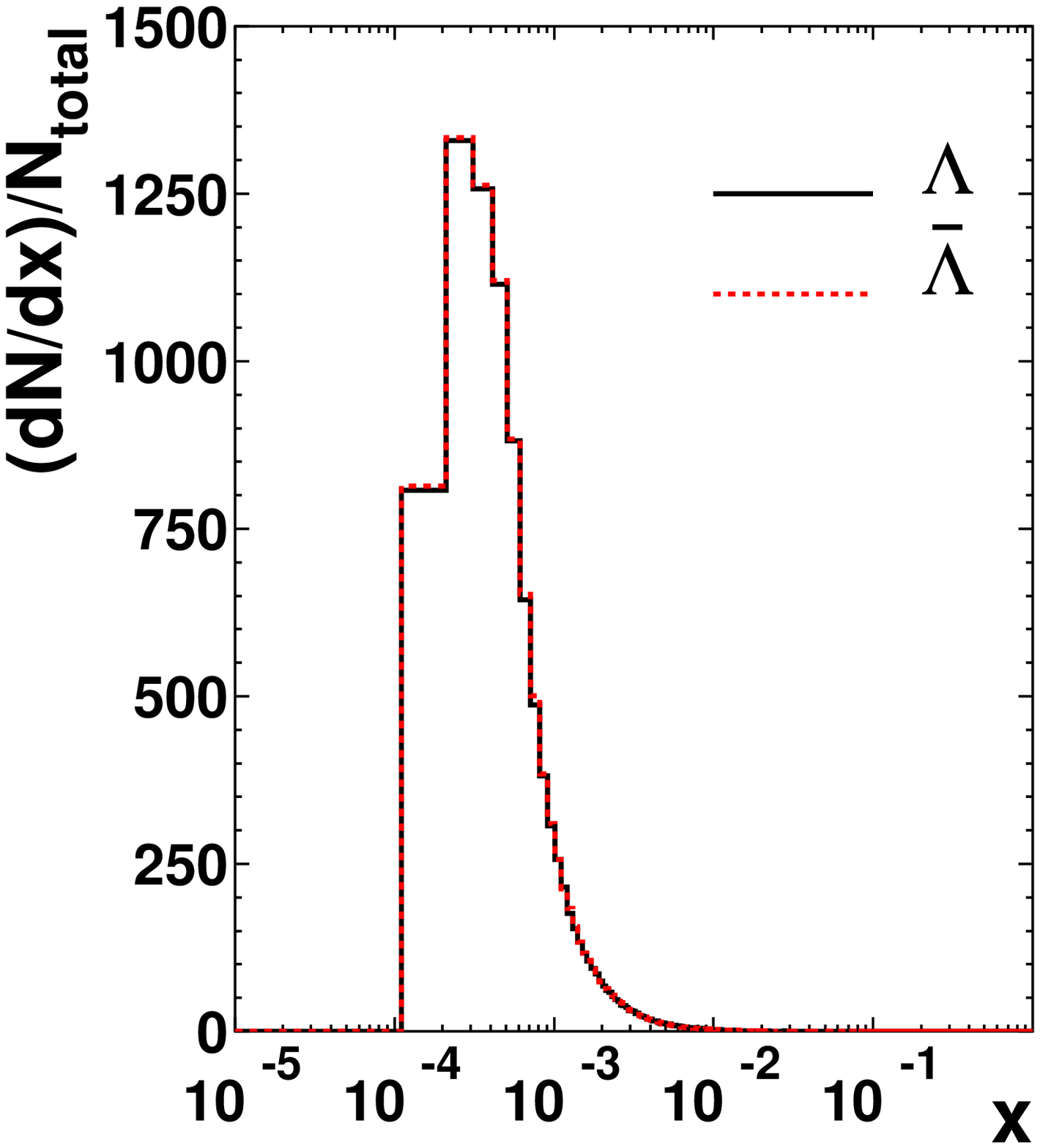}}
    \caption{The x-distribution of the struck quark or anti-quark that leads to the production of $\Lambda$ 
    or $\bar{\Lambda}$ in the kinematic region $x_F>0$ at eRHIC energy $\sqrt{s}$=100GeV. }
    \label{fig:xdeperhic}       
    \end{figure}

\begin{figure} 
    \resizebox{0.8\textwidth}{!}{\includegraphics{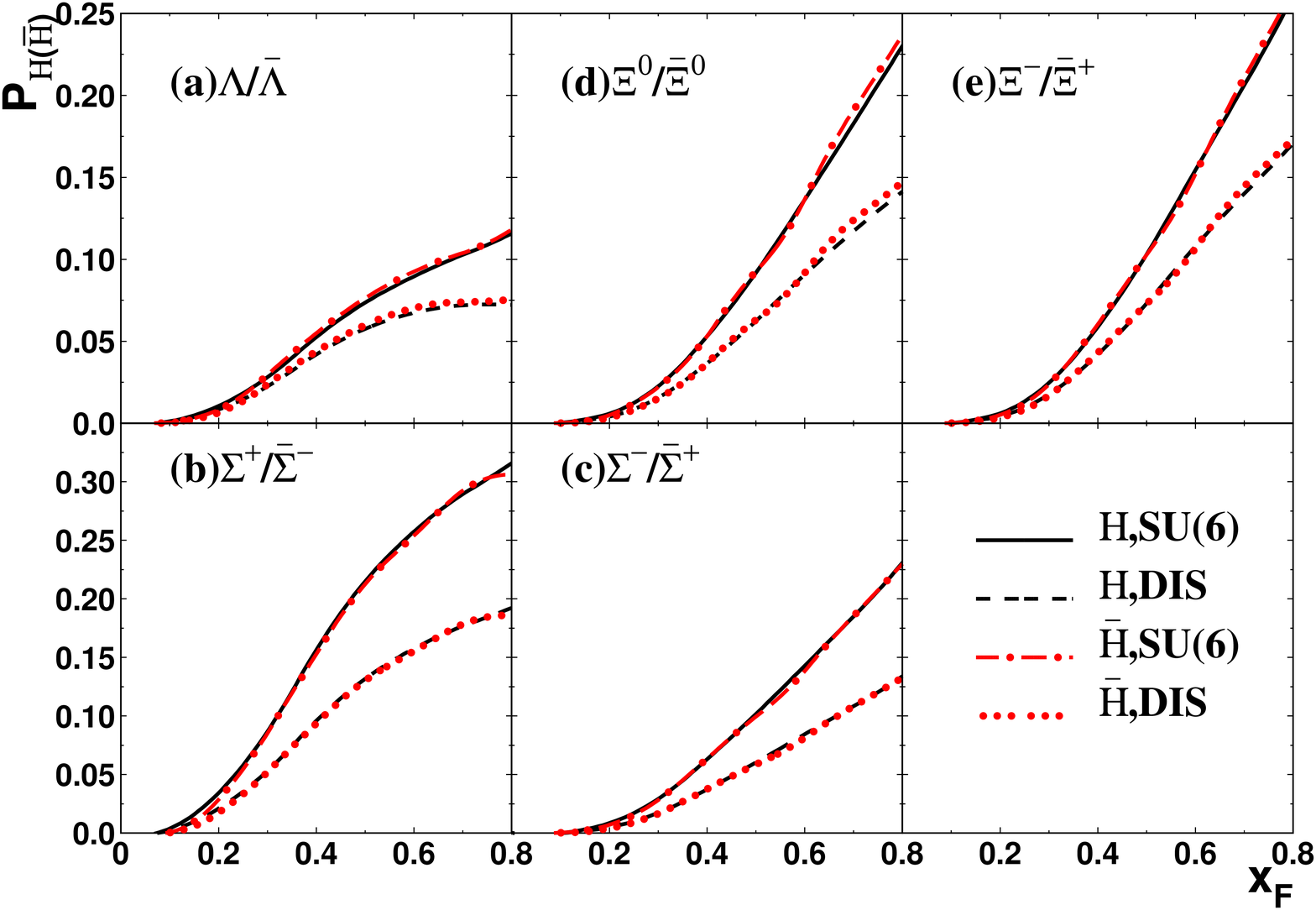}}
    \caption{(color online) Polarizations of hyperons and anti-hyperons versus $x_{F}$ at eRHIC energy
    $\sqrt{s}$=100.0 GeV obtained using a symmetric strange sea and anti-sea distribution.}
    \label{fig:polhyperhic}       
    \end{figure}

    It is also clear that, if we go to even  higher energies, the main contributions are from 
    even smaller $x$ region. In the small $x$ regions, valence quark contributions are negligible. 
    In such cases, we should have $P_H=P_{\bar H}$ assuming
    a symmetric sea and anti-sea quark distribution. 
    To show this, we made calculations at eRHIC energy, i.e., 
     we take the electron beam of 10 GeV and proton beam of 250 GeV.
    The electron beam polarization is taken as one and the nucleon is taken as unpolarized. 
    We first checked the $x$ distribution of the struck quarks (anti-quarks) that lead to the productions 
    of hyperons at such high energy. 
    In the calculations, we choose events in the kinematic region $0.2\le y\le 0.9$ and $Q^2>1$GeV$^2$. 
    The results are shown in Fig.~\ref{fig:xdeperhic}. 
    We see that they are indeed dominated by very small $x$. 
    The results of hyperon and anti-hyperon polarization using the same parton distribution
    set CTEQ2L with symmetric sea and anti-sea densities, are shown in
    Fig.~\ref{fig:polhyperhic}. 
    As expected, the $H$ and $\bar H$ polarization are almost the same. 
    
    \subsubsection{Results obtained using different sets of PDF's}
\begin{figure} [htb]
    \resizebox{0.4\textwidth}{!}{\includegraphics{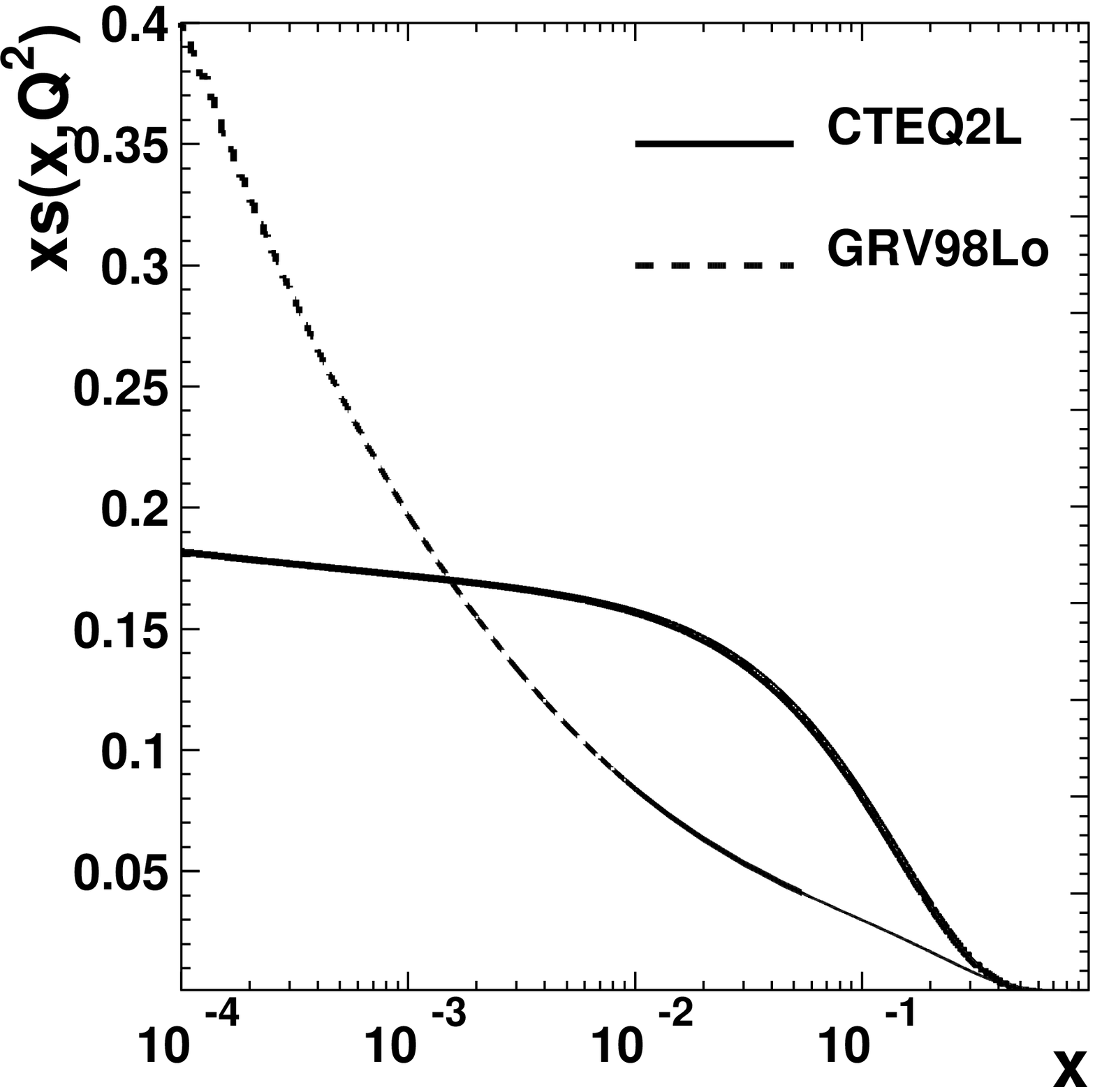}}
    \caption{Comparison of the sea quark distributions from GRV98Lo with those from CTEQ2L at $Q^2=3$GeV$^2$.}
    \label{fig:comparepdf}       
    \end{figure}

    In the calculations presented above, we used CTEQ2L for parton distributions. 
    As mentioned earlier, there are different sets of parameterizations available
    and the significant differences still exist for sea quark distributions especially for the strange sea.
    As an example, we show in Fig.~\ref{fig:comparepdf} the $s$ ($\bar s$) quark distribution 
    in CTEQ2L and GRV98Lo. 
    We see that the difference between the two parameterizations is indeed quite large. 
    
\begin{figure} [htb]
     \resizebox{0.6\textwidth}{!}{\includegraphics{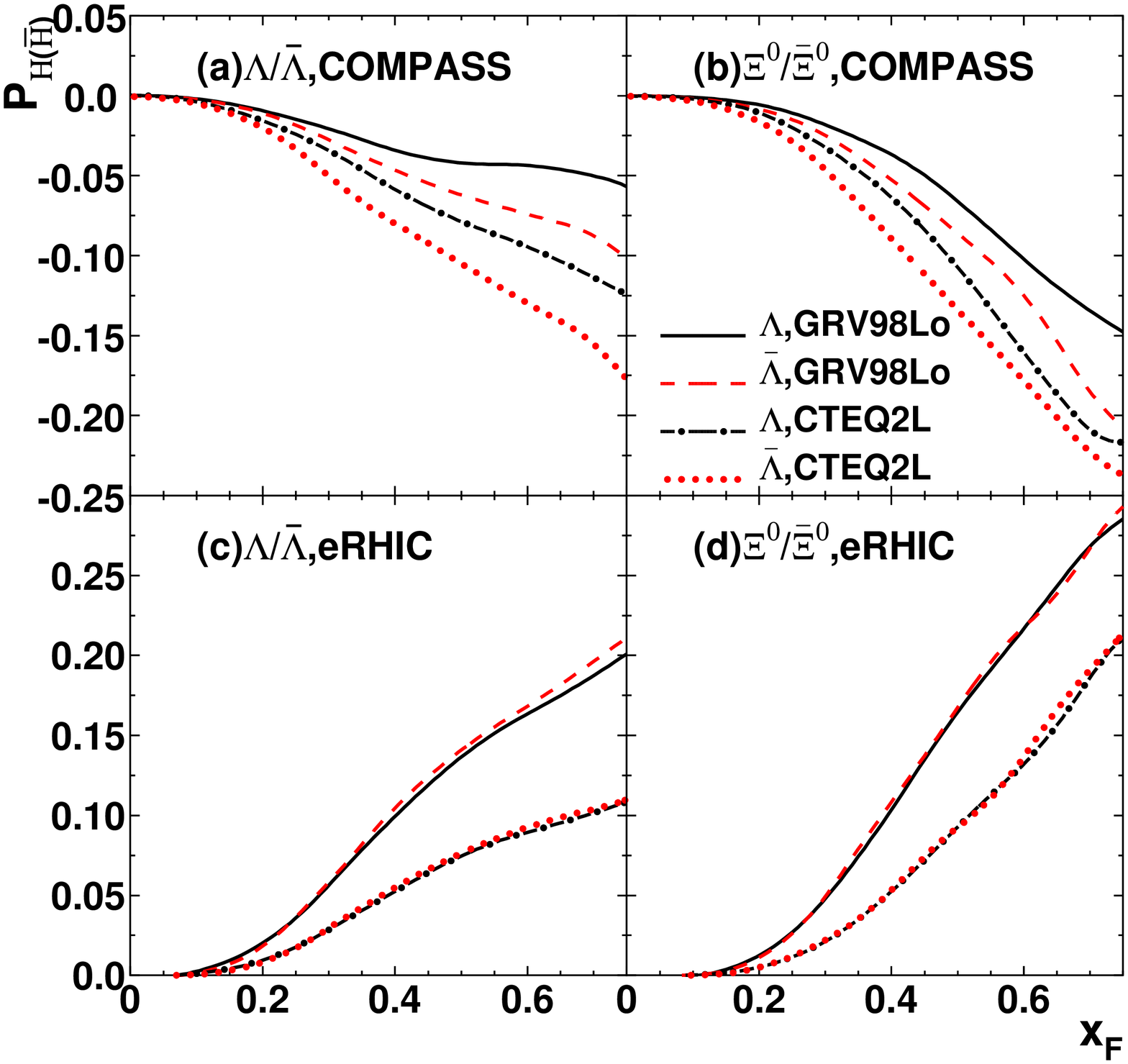}}
    \caption{(color online) Comparison of the polarizations of hyperons and anti-hyperons obtained using 
    GRV98Lo with those using CTEQ2L parton distribution functions 
    at the COMPASS energy (with beam polarization of $P_b=-0.76$) and 
    at the eRHIC energy (with $P_b=1$) respectively.
    For clarity, we only show the results obtained using  SU(6) picture for 
    spin transfer in fragmentation processes.}
    \label{fig:comparehyp}       
    \end{figure}

    The difference in different sets of PDF's can certainly influence the results of $P_H$ and $P_{\bar H}$.
    We study this influence by repeating the calculations mentioned above using different 
    sets of parton distribution functions. 
    As examples, we show the results for $\Lambda$, $\Xi^0$ and their anti-particles in
    Figs.~\ref{fig:comparehyp}(a) through (d) at COMPASS and eRHIC energies respectively.
     
   From the results, we indeed see some significant differences between the results obtained using 
   the two different sets of PDF's. 
   We see in particular that, at the COMPASS energy, the magnitude of the polarizations 
   obtained using CTEQ2L PDF's are larger than the corresponding results obtained 
   using GRV98Lo. 
   In contrast, at the eRHIC energy, the polarizations obtained using GRV98Lo PDF's are larger.
   This is because, at eRHIC energy, the dominating contributions are from very small $x$ region 
   where $s(x)$ in GRV98Lo is larger than that in CTEQ2L (see Fig.\ref{fig:comparepdf}). 
   However, at COMPASS energy, the dominating contributions are from much larger $x$ region, 
   where $s(x)$ in GRV98Lo is smaller than that in CTEQ2L.
    Such differences lead to different relative weights $R_f^H$ and manifest themselves 
    in the results for $P_H$ and $P_{\bar H}$ shown 
    in Figs.~\ref{fig:comparehyp}(a) through (d).
    We also see that different sets of PDF's influence the magnitudes of $P_H$ and $P_{\bar H}$ 
    but they have little influence on the difference between them. 
    The difference between $P_H$ and $P_{\bar H}$ is not very sensitive to the parameterizations of PDF's.

    \subsection{Results with asymmetric strange sea and anti-sea distribution}

\begin{figure}[htb] 
    \resizebox{0.5\textwidth}{!}{\includegraphics{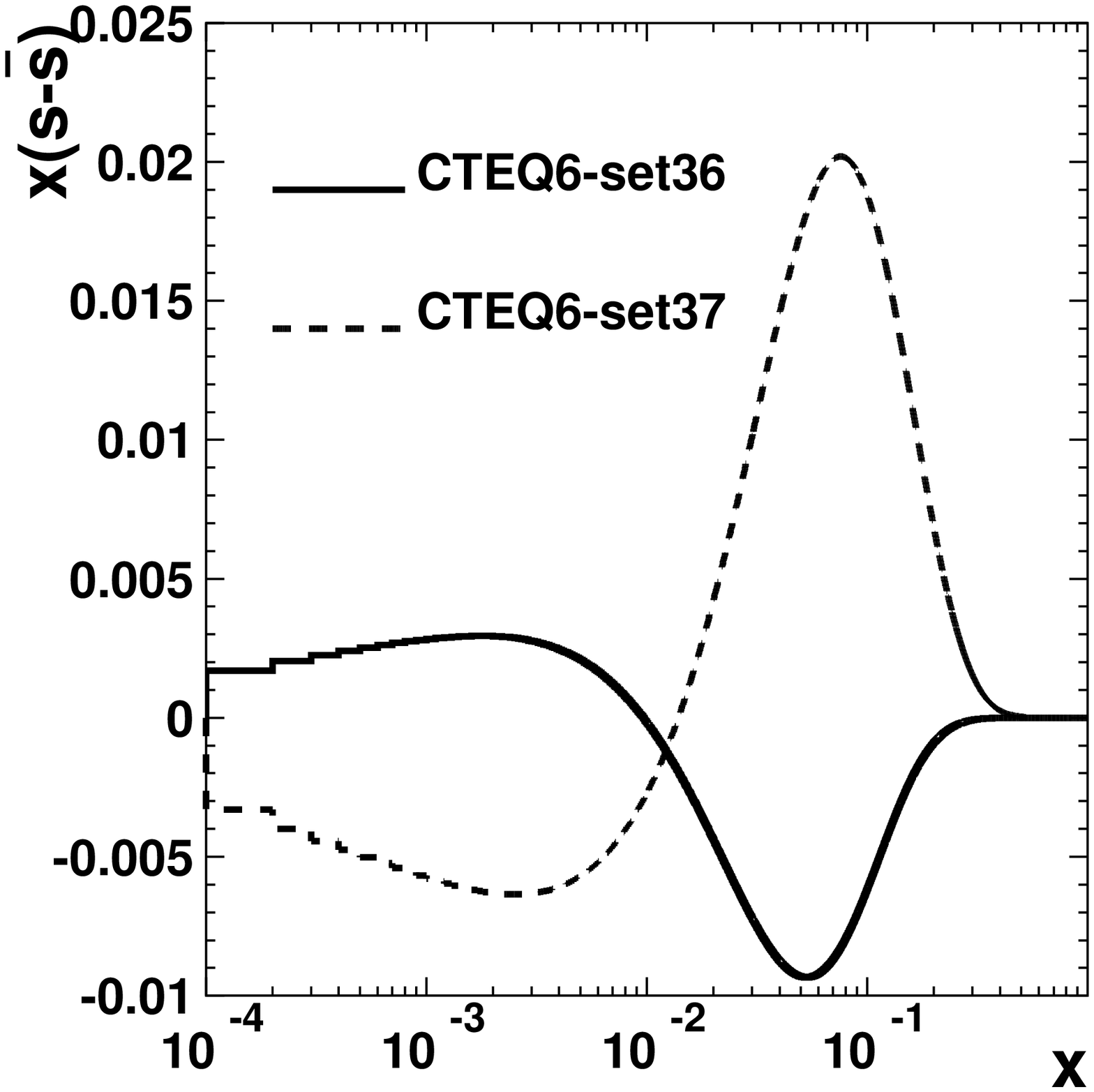}}
    \caption{Examples of the asymmetry of strangeness distributions in CTEQ6
    parameterizations at $Q^2=3$GeV$^2$. 
    We see in particular that they have opposite signs in most of the $x$ region.}
    \label{fig:asymmetry}       
    \end{figure}

    As discussed in last section, an asymmetry between strange sea and anti-sea quark
    distributions can be another source for the difference between hyperon and
    anti-hyperon polarization, and this effect remains at even higher energies such as at eRHIC. 
    The asymmetry in the strange sea of
    the nucleon was studied by many authors in 
    literature\cite{Signal:1987gz,Brodsky:1996hc,Burkardt:1991di,Holtmann:1996be,Christiansen:1998dz,Cao:1999da}.
    Different models are proposed. 
    A global QCD fit to the CCFR and NUTEV dimuon data has also shown a clear evidence 
    that $s(x)\neq \bar{s}(x)$ \cite{Olness:2003wz,Lai:2007dq}, 
    and a parameterization of the strangeness asymmetry has also been included in the
    CTEQ parameterization. 
    Such an asymmetry is usually described by defining $s^-(x)=s(x)-\bar s(x)$,  
    and correspondingly denote $s^{+}(x)=s(x)+\bar s(x)$.
    It seems now evident that $s^-(x)\not=0$ but the size is quite unknown besides 
    that it has to fulfill the limit $-s^+(x)\le s^{-}(x)\le s^+(x)$. 
    For example, we show two different parameterization from CETQ in Fig.~\ref{fig:asymmetry}.
    We see that the difference in the parameterization of $s^-(x)$ is indeed very large. 
    We even do not know the sign of  $s^-(x)$ in a given $x$-region.
    In this subsection, we study the contribution of such an asymmetry to the difference 
    between the polarization of $H$ and that of the corresponding $\bar H$ in SIDIS.
    
\begin{figure} 
    \resizebox{0.6\textwidth}{!}{\includegraphics{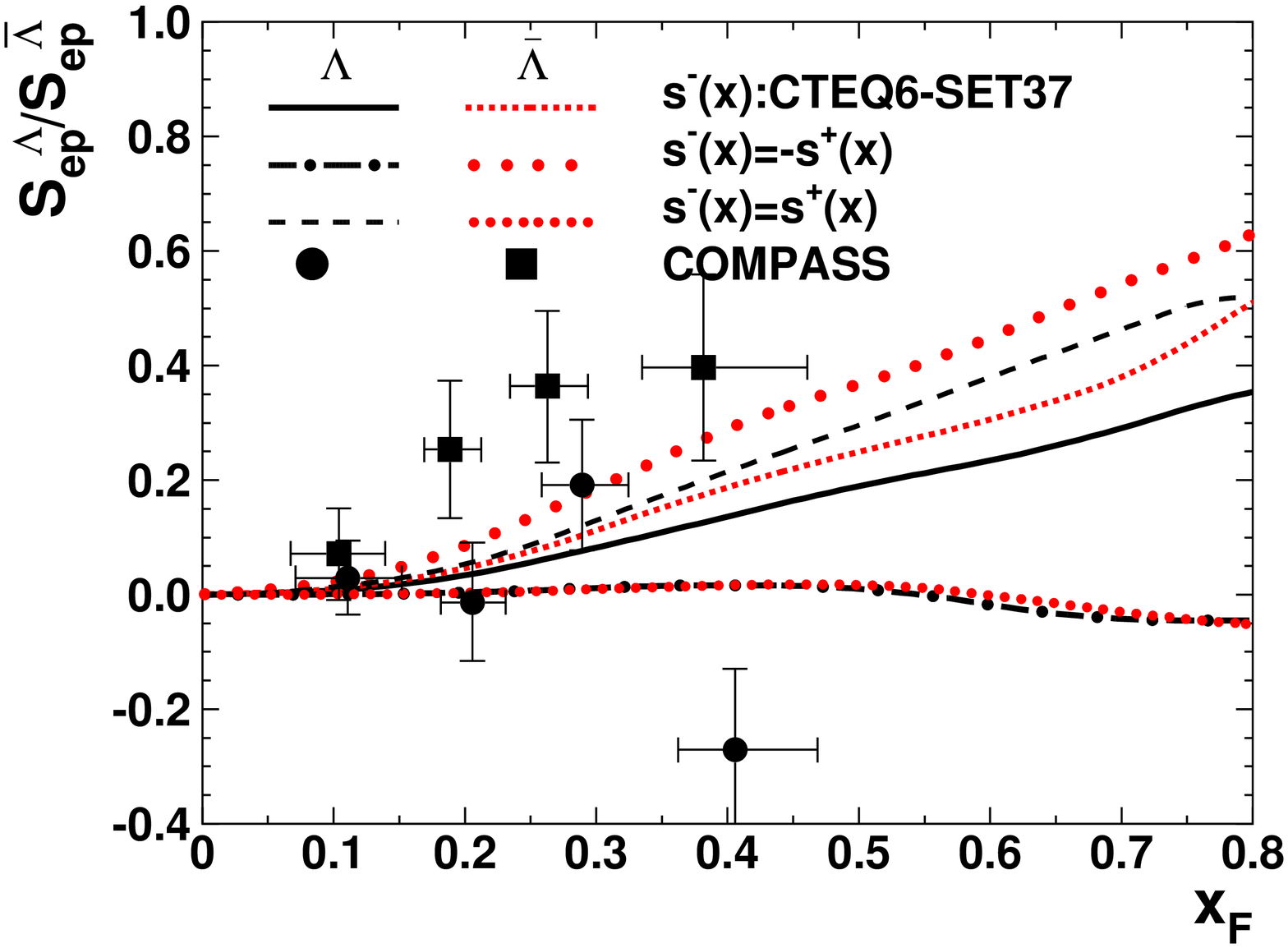}}
    \caption{(color online) Spin transfer $S_{ep}^{\Lambda}$ $S_{ep}^{\bar{\Lambda}}$ as functions
    of $x_{F}$ at the COMPASS energy $\sqrt{s}$=17.35 GeV obtained using 
    an asymmetric strangeness distribution in nucleon.
    Other PDF's are taken from CTEQ2L and SU(6) picture
    for spin transfer in fragmentation process are used.
    The data points are taken from COMPASS [7].}
    \label{fig:compasslamasy}       
    \end{figure}

    We first carried out the calculations by taking the same $s^+(x)$ and other PDF's from CTEQ2L 
    as used in previous calculations but taking a  $s^-(x)$ into account. 
    Since our knowledge of $s^-(x)$ is very much limited, and the form of $s^-(x)$ is almost completely unknown, 
    we simply take an existing parameterization as one from CETQ6set37.  
    With these inputs, we obtain the $\Lambda$ and $\bar\Lambda$ polarizations 
    and $S_{ep}^\Lambda$ and $S_{ep}^{\bar\Lambda}$ 
    in COMPASS kinematic region are obtained and are shown in Fig.~\ref{fig:compasslamasy}.
    We see that, in this kinematic region, the influence from such a small asymmetry $s^-(x)$ is small. 
    To see how large the effect can be, we take the extreme cases for $s^{-}(x)$, i.e.  $s^-(x)=-s^+(x)$ 
    or $s^-(x)=s^+(x)$. The results are also shown in Fig.~\ref{fig:compasslamasy}. 
    We see that the difference between $\Lambda$ and $\bar \Lambda$ in either limit is much larger
    than the case of symmetric $s(x)$ and $\bar s(x)$, and closer 
    the existing COMPASS data.\cite{Alexakhin:2005dz,Sapozhnikov:2005sc}
    
\begin{figure} 
    \resizebox{0.4\textwidth}{!}{\includegraphics{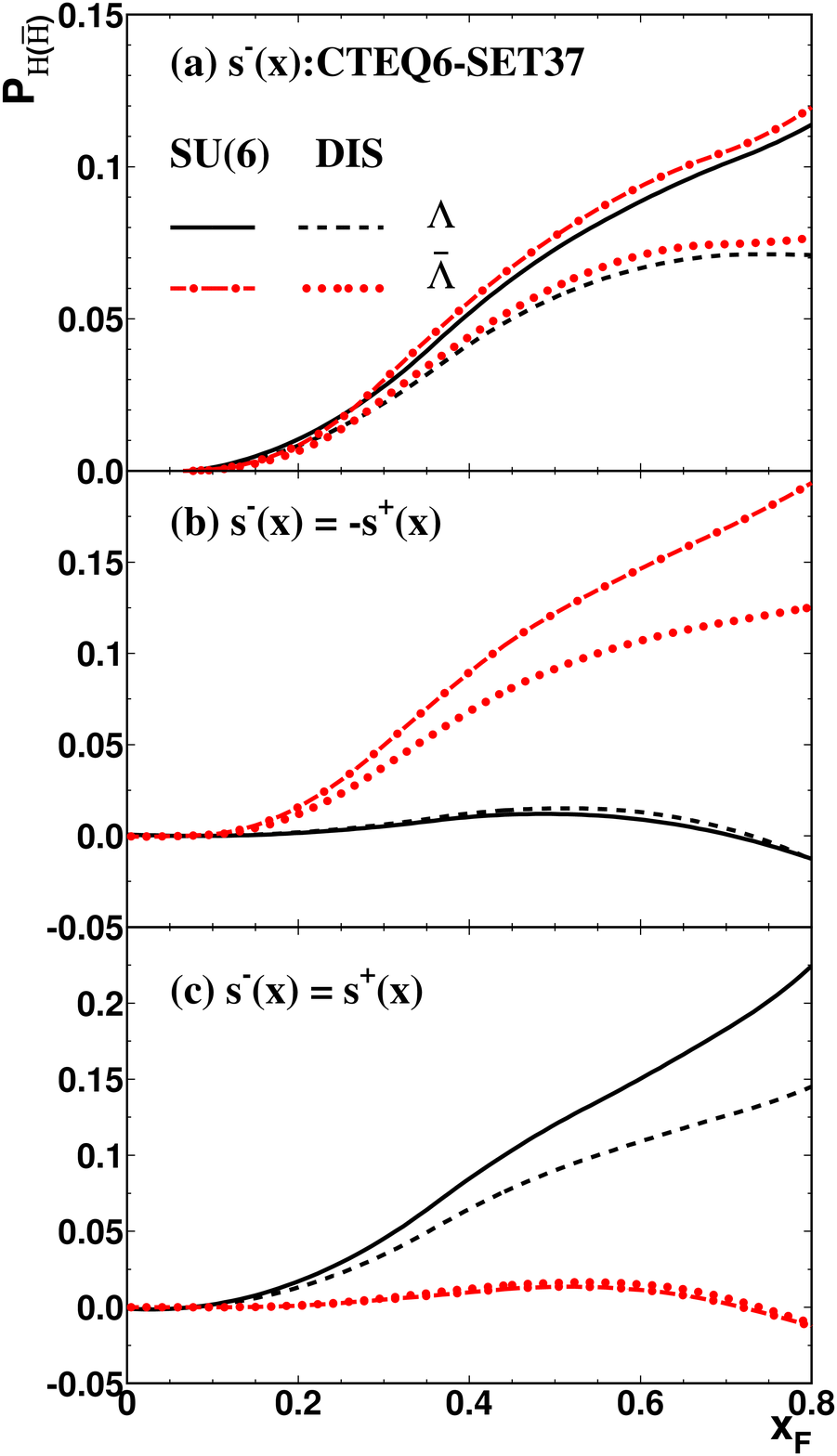}}
     \resizebox{0.4\textwidth}{!}{\includegraphics{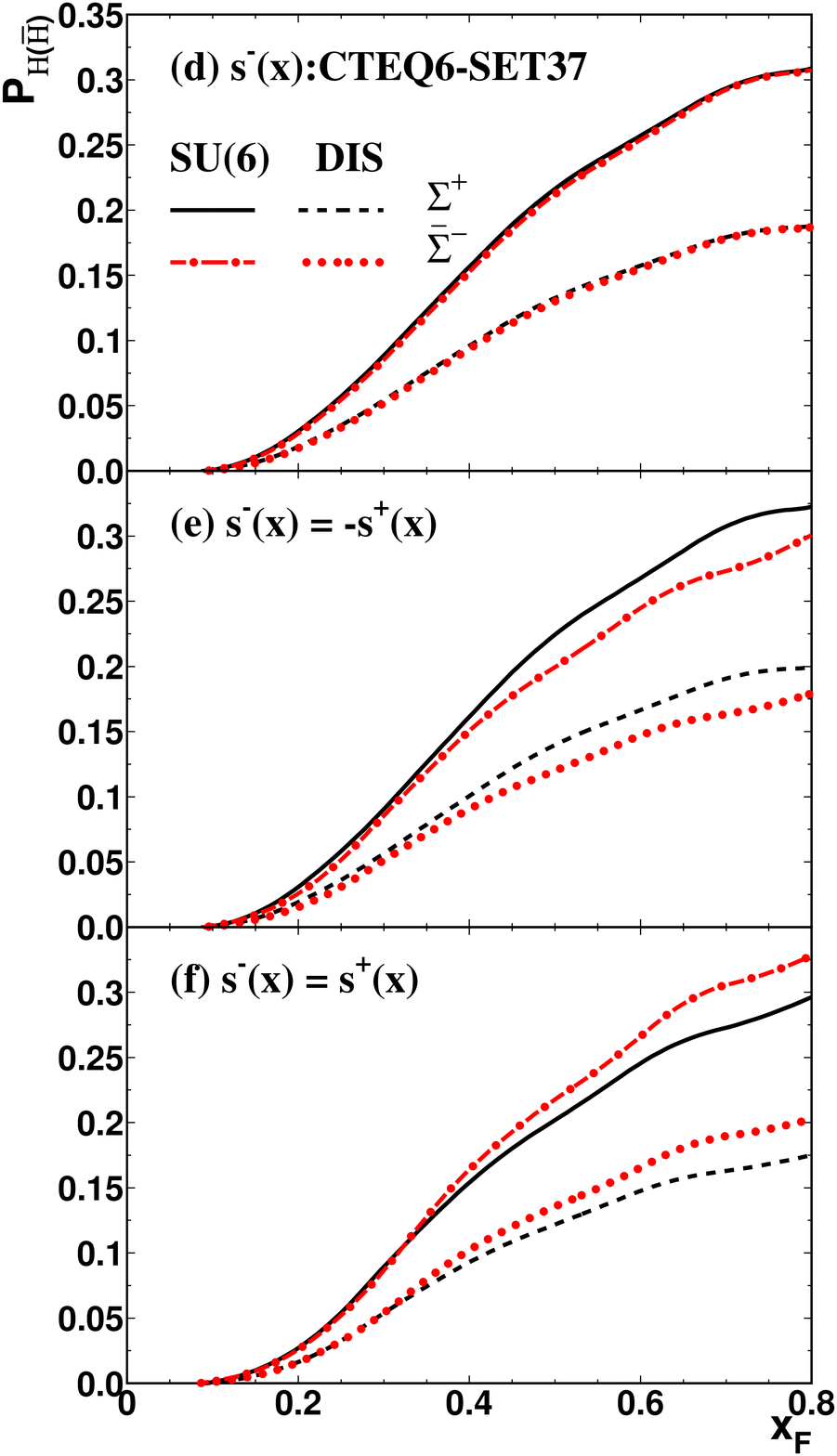}}
     \caption{(color online) Longitudinal polarizations of $\Lambda$, $\Sigma^+$ and their anti-particles 
    obtained using different asymmetric strangeness distributions in nucleon at eRHIC energy.}
    \label{fig:erhicPolasy}       
    \end{figure}

    At the eRHIC energy, the only source for the difference between $P_{\Lambda}$ and $P_{\bar{\Lambda}}$ 
    is the asymmetry between $s(x)$ and $\bar s(x)$.
    We did similar calculations and obtain the results shown in Fig.~\ref{fig:erhicPolasy}.
    We can see that the difference between $P_{{\Lambda}}$ and $P_{\bar{\Lambda}}$ 
    is quite small if we use the asymmetric strangeness distribution as given in CTEQ6set37.
    However, it can be rather large at the extreme case. 
    The asymmetry between $s(x)$ and $\bar s(x)$ has even smaller influence on $P_\Sigma$ 
    and their anti-particle.
    For comparison, we show the results for $\Sigma^+$ and $\bar\Sigma^-$ in the same figure. 
    These results show us that, experiments at eRHIC can indeed provide us useful information 
    on the asymmetry between $s(x)$ and $\bar s(x)$ in nucleon, but high statistics is needed. 
    
    \section{$P_H$ and $P_{\bar H}$ in SIDIS with unpolarized beam and polarized target}

\begin{figure} 
    \resizebox{0.5\textwidth}{!}{\includegraphics{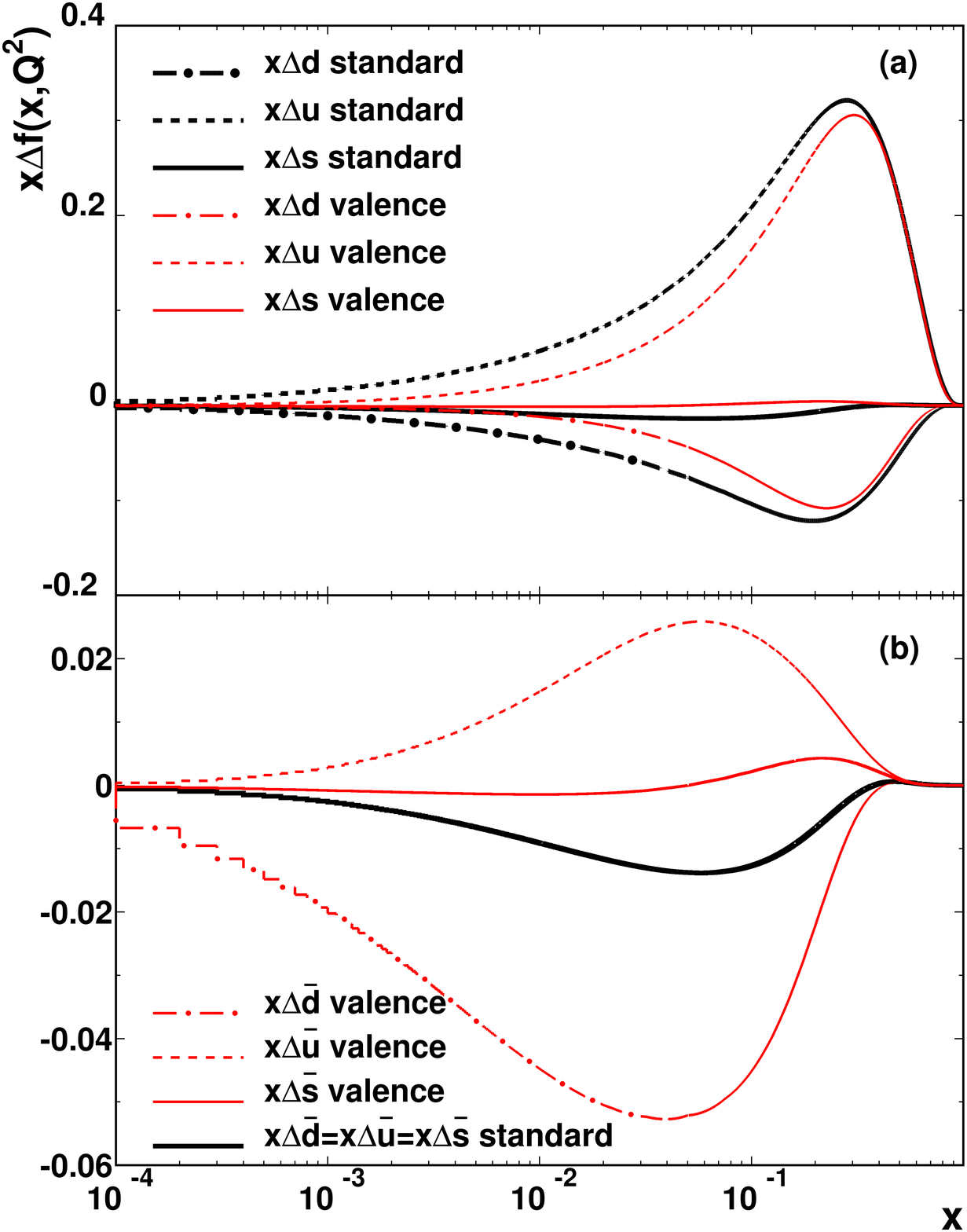}}
    \caption{(color online) Comparison of the polarized quarks (a) and anti-quarks (b) distributions obtained from leading
    order GRSV2000 standard and valence scenario at $Q^2= 3$GeV$^2$.}
    \label{fig:grsvsea}       
    \end{figure}

\begin{figure}[htb] 
     \resizebox{0.6\textwidth}{!}{\includegraphics{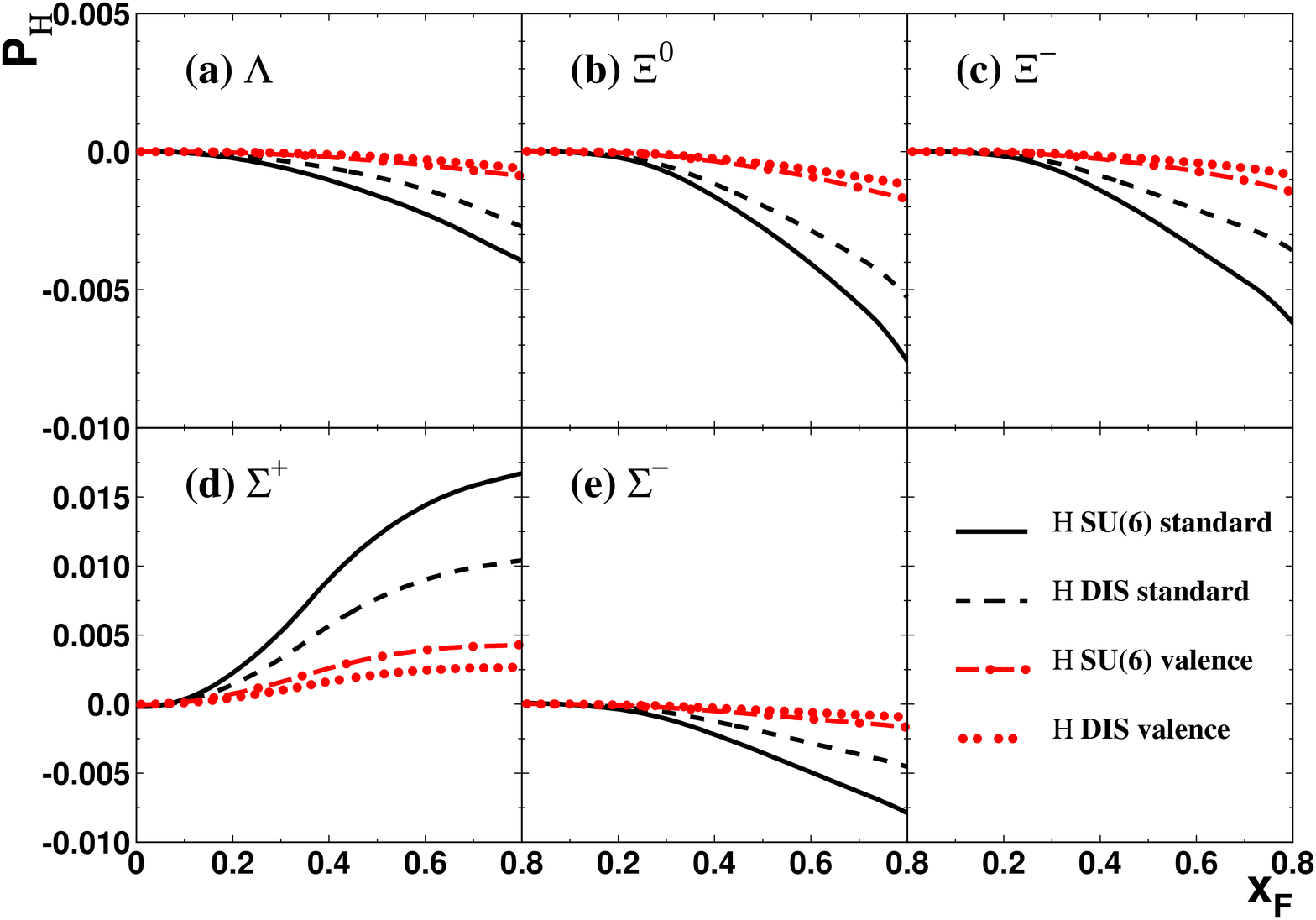}}
    \resizebox{0.6\textwidth}{!}{\includegraphics{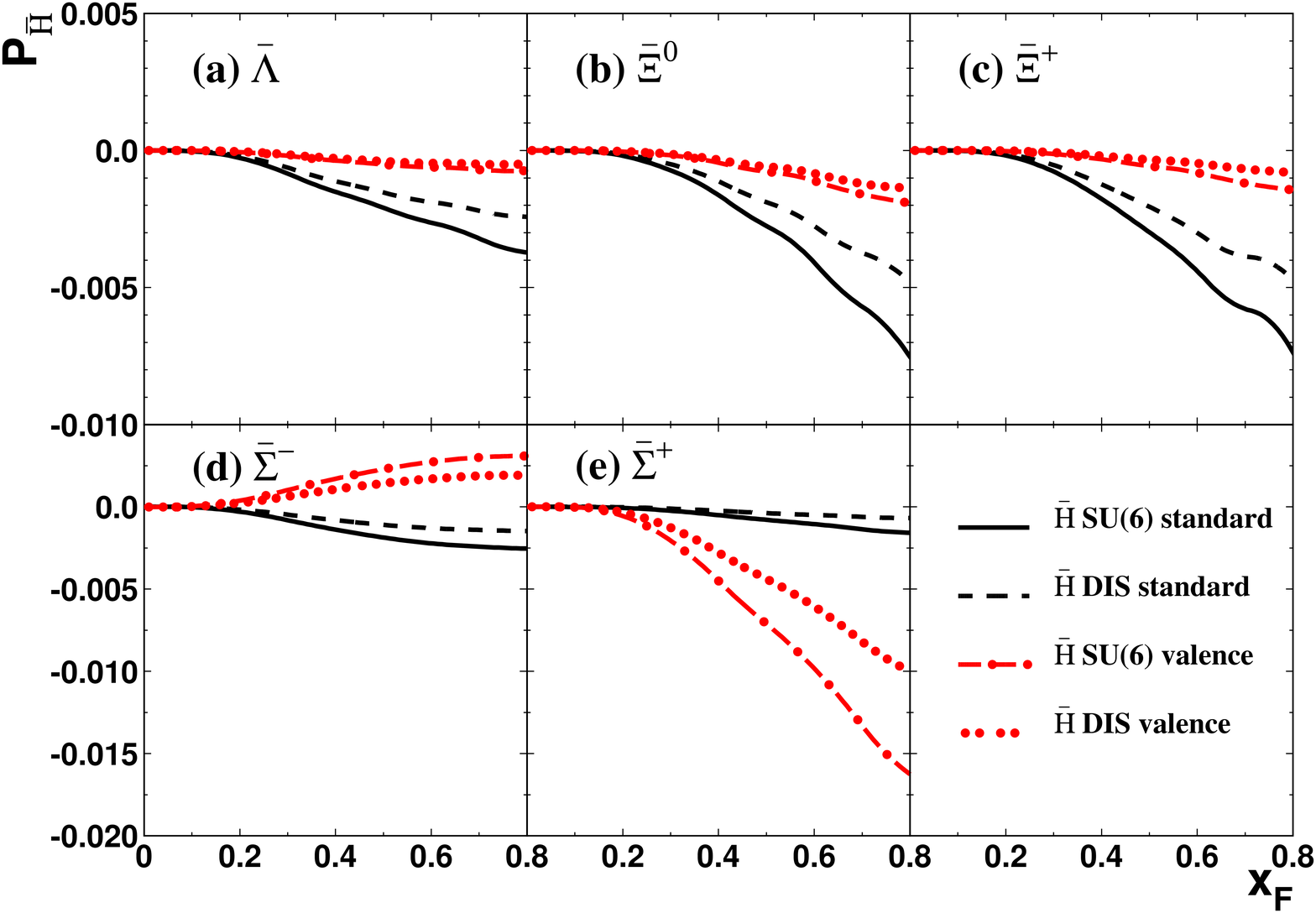}}
    \caption{(color online) Longitudinal polarizations of the hyperons(upper) and anti-hyperons (lower) as a function of $x_{F}$ 
    at eRHIC energy with the longitudinal polarized target. } 
    \label{fig:erhicpol}       
    \end{figure}

    If the lepton beam is unpolarized and the target proton is polarized with $P_{T}=1$, 
    the polarizations of the hyperons (or anti-hyperons) are determined by Eq.~(\ref{eq:polH|Pb=0}). 
    In this case,  the relative weights for the contributions of different flavors are the same as 
    those discussed in last section which are determined by the unpolarized quantities. 
    However, the polarizations of the quarks and anti-quarks are different.  
    In the case of unpolarized lepton beam and longitudinally polarized nucleon, 
    and the polarizations of the quarks and anti-quarks equal to those 
    in the polarized nucleon,  which is a simple result of helicity conservation. 
    This is a good place to study polarized quark distributions in the nucleon. 
    There exist many different sets of parameterizations 
    of the polarized PDF's [see e.g. \cite{Gluck:1995yr,Bluemlein:2002be,Leader:2005ci,
    Gehrmann:1995ag,Hirai:2003pm,deFlorian:2000bm}]
    and the differences between them are quite large.  
    An example is given in Fig.~\ref{fig:grsvsea} where two sets of parameterizations 
    from GRSV2000\cite{Gluck:1995yr}, GRSV2000 set3 (standard) and set4 (valence),  are shown.
    We make calculations of $P_H$ and $P_{\bar H}$ using these two sets of parameterizations 
    of the polarized PDF's to see the sensitivity of the results of $P_H$ and $P_{\bar H}$ on the polarized PDF's.
    We carried out the calculations in the COMPASS kinematic region and at eRHIC energy. 
    The results at the two energies are similar and those at COMPASS energy 
    are shown in Ref.~\cite{Dong:2005ea}. 
    We show those at eRHIC energy in Fig.~\ref{fig:erhicpol}. 
    
    The results show in particular following interesting features. 
    First,  the polarizations of hyperons and anti-hyperons are quite sensitive to the polarized 
    PDF's.  Different sets of polarized PDF's lead indeed to 
    quite different results of hyperon and anti-hyperon polarizations.  
    We see in particular that the differences obtained from different set of polarized PDF's 
    are generally larger than the differences between the
    results for different models for the spin transfer in fragmentation.
    Second, because the relative weights $R_f^H$ and spin transfer $S_f^H$ are 
    quite different from each other for different flavor $f$ for a given hyperon $H$, 
    the polarizations of different hyperons and anti-hyperons are sensitive to
    polarized PDF's of different flavors. 
    For example, $P_{\Sigma^{+}}$ and $P_{\Sigma^{-}}$ are sensitive to 
    $\Delta{u(x)}$ and  $\Delta{d(x)}$ respectively. 
    They have different signs because the sign of $\Delta{u(x)}$ is different from that of $\Delta{d(x)}$. 
    The magnitude of  $P_{\Sigma^{+}}$ is larger than $P_{\Sigma^{-}}$ because 
    $|\Delta{u(x)}|$$>$$|\Delta{d(x)}|$.
    Similar features can be seen for $\Xi^{0}$, $\Xi^{-}$ and the corresponding anti-hyperons. 
     These two features are important because they show that we can use hyperon polarizations 
    in SIDIS to extract information on polarized PDF's. 
   
    \section{Summary and outlook}
    In summary, we have calculated the longitudinal polarizations of the
    hyperons and anti-hyperons in semi-inclusive deep-inelastic
    scattering at COMPASS and eRHIC energies. 
    We have in particular made a systematic study of the different contributions 
    to the differences between the polarization of a hyperon and its anti-particle. 
    We presented the results obtained in SIDIS with polarized beam and unpolarized target 
    for the case that a symmetric strange sea 
    and anti-sea distribution is used and those obtained in the case that 
    an asymmetry between  strange sea and anti-sea distribution is taken into account 
    and for reactions with unpolarized beam and polarized target.
    Our results show that, (1) at COMPASS energy, valence contributions play 
    an important role in the difference between hyperon and anti-hyperon polarization 
    but are negligible at eRHIC energy; 
    (2) a significant asymmetry between strange sea and anti-sea distributions 
    can manifest itself in the difference between hyperon and anti-hyperon polarization 
    at eRHIC energy, but high statistics is needed in order to detect it; 
    (3) different sets of PDF parameterizations have quite large influence on the magnitudes 
    of hyperon polarizations but the influence on the difference between 
    hyperon and anti-hyperon polarization is relatively small; 
    (4) hyperon and anti-hyperon polarizations in reactions using unpolarized beam 
    and polarized target are sensitive to the polarized parton distributions and different 
    hyperons are sensitive to different flavors, and hence can be used to extract information on 
    flavor tagging. 
   These results show that both the difference between hyperon and anti-hyperon polarization 
   in reaction with polarized beam and unpolarized target and the polarizations 
   of hyperons and anti-hyperons in reactions with unpolarized beam and polarized target 
   are sensitive to the sea structure of nucleon. 
    High precision measurements in particular those at high energies such as at 
    eRHIC are able to provide us deep insights into the nucleon sea.
    
    This work was supported in part by the National Natural Science Foundation of China 
    under the approval No. 10525523 and Department of Science and Technology of 
    Shandong Province.

    \end{document}